\documentclass[journal]{IEEEtran}

\usepackage{xcolor,soul,framed}

\ifCLASSINFOpdf
   \usepackage[pdftex]{graphicx}
   \graphicspath{{../pdf/}{../jpeg/}}
   \DeclareGraphicsExtensions{.pdf,.jpeg,.png}
\else
   \usepackage[dvips]{graphicx}
   \graphicspath{{../eps/}}
   \DeclareGraphicsExtensions{.eps}
\fi
\usepackage{amsmath}
\usepackage{
algorithm,
algpseudocode
}

\usepackage{booktabs}

\ifCLASSOPTIONcompsoc
 \usepackage[caption=false,font=normalsize,labelfont=sf,textfont=sf]{subfig}
\else
 \usepackage[caption=false,font=footnotesize]{subfig}
\fi
\begin{document}
%
\title{Disaggregating Customer-level Behind-the-Meter PV Generation Using Smart Meter Data and Solar Exemplars}
%
%
%


\author{Fankun Bu,~\IEEEmembership{Graduate Student Member,~IEEE,}
        Kaveh Dehghanpour,~\IEEEmembership{}
        ~Yuxuan Yuan,~\IEEEmembership{Graduate Student Member,~IEEE,}
        ~Zhaoyu Wang,~\IEEEmembership{Senior Member,~IEEE,}
        and Yifei Guo,~\IEEEmembership{Member,~IEEE}
\thanks{This work was supported in part by the National Science Foundation under EPCN 2042314 and in part by the Advanced Grid Modeling Program at the U.S. Department of Energy Office of Electricity under Grant DE-OE0000875. (\textit{Corresponding author: Zhaoyu Wang})}
\thanks{F. Bu, K. Dehghanpour, Y. Yuan, Z. Wang and Y. Guo are with the Department of
Electrical and Computer Engineering, Iowa State University, Ames, IA 50011
USA (e-mail: fbu@iastate.edu; wzy@iastate.edu).}
}

%
%

\markboth{}%
{Shell \MakeLowercase{\textit{et al.}}: Bare Demo of IEEEtran.cls for IEEE Journals}
%



\maketitle

\begin{abstract}
Customer-level rooftop photovoltaic (PV) has been widely integrated into distribution systems. In most cases, PVs are installed behind-the-meter (BTM), and only the net demand is recorded. Therefore, the native demand and PV generation are unknown to utilities. Separating native demand and solar generation from net demand is critical for improving grid-edge observability. In this paper, a novel approach is proposed for disaggregating \textit{customer-level} BTM PV generation using low-resolution but widely available hourly smart meter data. The proposed approach exploits the strong correlation between monthly nocturnal and diurnal native demands and the high similarity among PV generation profiles. First, a joint probability density function (PDF) of monthly nocturnal and diurnal \textit{native} demands is constructed for customers \textit{without} PVs, using Gaussian mixture modeling (GMM). Deviation from the constructed PDF is utilized to probabilistically assess the monthly solar generation of customers \textit{with} PVs. Then, to identify \textit{hourly} BTM solar generation for these customers, their estimated monthly solar generation is decomposed into an hourly timescale; to do this, we have proposed a maximum likelihood estimation (MLE)-based technique that utilizes hourly typical solar exemplars. Leveraging the strong monthly native demand correlation and high PV generation similarity enhances our approach's robustness against the volatility of customers' hourly load and enables highly-accurate disaggregation. The proposed approach has been verified using real native demand and PV generation data.
\end{abstract}


%
\IEEEpeerreviewmaketitle

\section{Introduction} \label{sec:intro}
%
%
%
%

\IEEEPARstart{I}{N} practice, customer-level rooftop PVs are integrated into distribution systems at behind-the-meter (BTM), where only the net demand is recorded. The measured net demand equals native demand minus the PV generation, which are unknown to utilities separately. The native demand refers to the original demand consumed by home appliances. The invisibility of native demand and BTM solar generation poses challenges in distribution network design \cite{hosting_capacity,ying_zhang}, operation and expansion \cite{PV_handbook,restoration,PV_increasing}, load/PV generation forecasting \cite{Yi_Wang, kangping_recom_1}, and demand response \cite{kangping_recom,kangping_li_solar}. Thus, disaggregating PV generation from net demand is of significance to utilities.

Previous works regarding PV generation disaggregation can be classified into two categories based on the scale of solar power: \textit{Class I - Customer-level approaches:} Customer-level BTM PV generation disaggregation can provide more fine-grained spatial granularity to utilities. Thus, the separated PV generation and native demand for individual customers can be aggregated to obtain the estimate at any higher levels, i.e., service transformer, feeder, or substation. In \cite{SunDance}, customer PV generation is estimated by combining a PV performance model with a clear sky model, and using meteorological/geographical data. In \cite{applied_energy}, a non-intrusive load monitoring (NILM) approach is proposed to disaggregate customers' PV generation from their net demand using measurements with 1-second resolution. In \cite{kangping_li_solar,PV_cap_estimation}, a data-driven method is proposed for estimating the capacity and power output of residential rooftop PVs using customers' net load curve features. In \cite{YC_Zhang, nan_peng_yu_2}, a physical PV performance model is combined with a statistical load estimation model, along with weather data to identify key PV array parameters. The disadvantages of previous customer-level approaches are as follows: dependency on the availability of accurate native demand exemplars, unavailability of PV model parameters, requiring high-resolution sensors and weather data. These obstacles make the previous methods susceptible to the uncertainties of customer behavior and rooftop solar power generators, which result in a decline in disaggregation accuracy. 

\textit{Class II - System-level approaches:} Many previous works have proposed methods to disaggregate solar power from net demand at transformer, feeder, or regional levels. In \cite{indu_informatics}, a data-driven approach is presented for separating the aggregate solar power of groups of customers using their service transformer measurements. In \cite{CSSS_feeder}, an exemplar-based disaggregator is proposed to separate the output power of an unobservable solar farm from the feeder-level $\mu$PMU measurements, using power measurements of nearby observable PV plants and irradiance data. In \cite{Yi_Wang}, a regional-scale equivalent PV station model is proposed to represent the total generation of small-scale PVs. The model parameters are optimized using known solar power data. In \cite{Hamid_Shaker}, a data-driven approach is proposed to estimate the total rooftop PV generation in a region by installing temporary sensors to measure representative solar arrays. Furthermore, previously in \cite{Fankun_Bu}, we developed a game-theoretic data-driven approach for disaggregating the PV generation of sizeable groups of customers using solar and load exemplars. However, Class II approaches lack sufficient accuracy for performing customer-level PV disaggregation. 

Considering the shortcomings of previous approaches, we propose a novel customer-level solar power disaggregation technique. Our basic idea is to first estimate each customer’s \textit{monthly} BTM PV generation and then decompose it into hourly solar power using solar exemplars. Note that in geographically bounded distribution systems, solar exemplars can be easily constructed from observable PVs due to the strong spatial correlation in weather data. Merely having solar exemplars is not sufficient to estimate the unknown PV generation; the relationship between the solar exemplar and unknown PV generation needs to be identified. One promising solution is to construct native demand exemplars. However, accurate customer-level native demand exemplar at the hourly timescale cannot be obtained due to high load uncertainties. To tackle this problem, we exploit an observation from our real smart meter data that the monthly nocturnal and diurnal native demands are highly correlated. Note that this high correlation applies to customers both with and without PVs. Then, identifying the relationship between the solar exemplar and unknown PV generation comes down to making the known monthly nocturnal native demand and the estimated monthly \textit{diurnal} native demand optimally conform to the observed correlation. In other words, to avoid directly identifying the relationship at the hourly timescale, we first identify it at the monthly timescale and then extend the identified relationship to the hourly timescale.

More specifically, the first step is to construct the joint probability density function (PDF) of monthly nocturnal and diurnal native demands for \textbf{\textit{customers without PVs}}. This will be done using a Gaussian Mixture Model (GMM) technique \cite{pattern_recog_book}, which has demonstrated significant flexibility in forming smooth approximations to arbitrarily-shaped PDFs. The constructed joint PDF captures the monthly load characteristics of customers without PVs; hence, this joint PDF serves as a benchmark for evaluating the deviations caused by monthly BTM solar generation for \textit{\textbf{customers with unobservable PVs}}. The second step is to project the obtained customer-level monthly solar estimations onto hourly values; to do this, the monthly BTM solar generations are represented as a linear weighted summation of solar exemplars with hourly resolution. The weights are optimized using a constrained maximum likelihood estimation (MLE) process, and will be leveraged to disaggregate the hourly net demand of customers with BTM PV generators. To enhance the robustness of MLE against anomalous data, a penalty term is integrated into the weight identification process. Throughout the paper, vectors are denoted using bold italic letters, and matrices are denoted as bold non-italic letters.

The main contributions of our paper are summarized as follows: (1) Our approach takes full advantage of the strong similarity among small-scale rooftop PV generations. This similarity is due to the fact that the PVs installed within a spatially-bounded distribution system are subject to nearly identical meteorological inputs. (2) The proposed technique utilizes the significant correlation between monthly nocturnal and diurnal native demands. In this way, our approach avoids the direct use of hourly native demand, which is highly volatile at the customer level \cite{PV_load_uncertainty, load_uncertain}.
(3) Our approach innovatively leverages a soft margin to mitigate the impact of anomalous data samples of solar exemplars. The introduction of this penalty term enhances the robustness of our approach against abnormal measurements.

The rest of the paper is organized as follows: Section \ref{sec:overall} introduces the overall framework for customer-level BTM PV generation disaggregation and describes smart meter dataset. Section \ref{sec:GMM_construction} presents the process for constructing joint PDF of monthly diurnal and nocturnal native demands. Section \ref{sec:MLE} describes the procedure of formulating and solving MLE to perform disaggregation. In Section \ref{sec:casestudy}, case studies are analyzed. Section \ref{sec:discussion} discusses the relevant applications of the disaggregated estimates and Section \ref{sec:conclusion} concludes the paper.

\begin{figure}[t]
\centering
\includegraphics[width=0.75\linewidth]{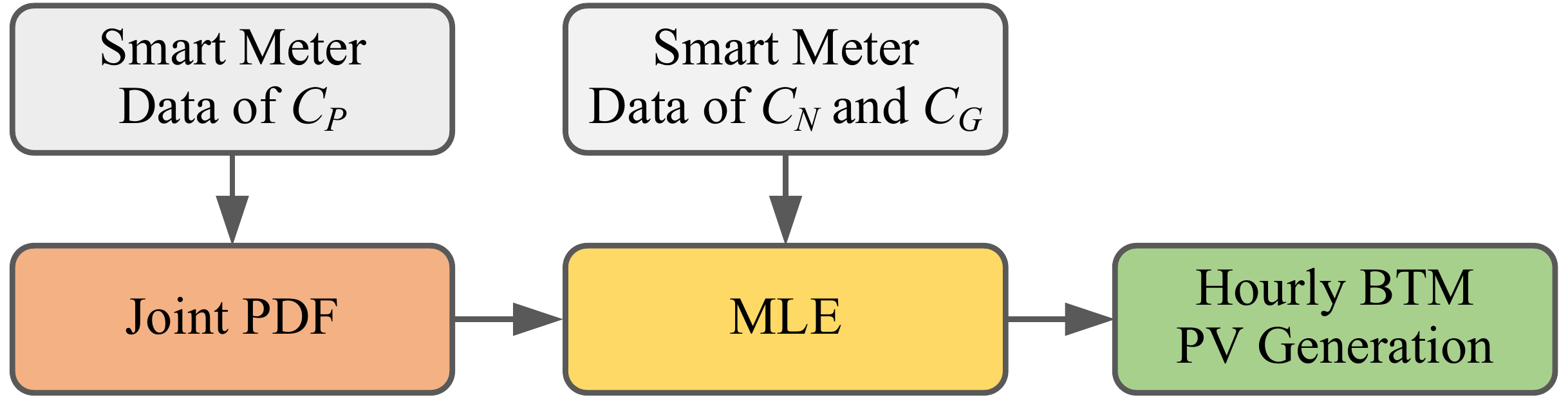}
\caption{Overall structure of the proposed customer-level BTM PV generation disaggregation method.}
\label{fig:flow_chart}
\end{figure}

\section{Overall Disaggregation Framework and Dataset Description}\label{sec:overall}
\subsection{Overall Framework}
In distribution systems, residential customers can be typically categorized into three types: (I) $C_P$ is the set of customers \textit{without} PVs whose native demand is recorded by smart meters. (II) $C_G$ denotes the small group of customers \textit{with} PVs whose PV generation and native demand are both observable separately. (III) $C_N$ represents the set of customers with PVs whose \textit{net demand} is recorded by smart meter, while their native demand and PV generation are not separately visible. Our goal is to disaggregate PV generation and native demand from the net demand of individual customers in $C_N$.

The overall process is illustrated in Fig. \ref{fig:flow_chart}: First, the known monthly nocturnal and diurnal native demands of customers in $C_P$ are employed to construct a joint PDF using GMM modeling technique. This joint PDF is constructed based on a sizeable number of customers without PVs. Then, for each customer in $C_N$, the unknown PV generation is optimally estimated by performing MLE, and using the constructed joint PDF, known monthly net demand and solar exemplars. 

\subsection{Dataset Description}
The hourly native demand data used in this paper are from a Midwest U.S. utility \cite{Test_system}, and the hourly PV generation data are from a public dataset \cite{data_source}. The time range of solar power is one year, and the time range of native demand of customers without PVs is three years. The test system consists of 1120 customers, of which 480 are residential customers without PVs and 237 are residential customers with PVs. Net demand data is obtained by aggregating customers' PV generation and native demand data.

\section{Statistical Modeling of Monthly Native Demand}\label{sec:GMM_construction} 
\subsection{Findings from Real Smart Meter Data}\label{sec:Aggregation} 

\begin{figure}[htbp]
\centering
\subfloat[Customers w/o PV (Hourly)\label{sfig:corr_demand_hr}]{
\includegraphics[width=0.41\linewidth]{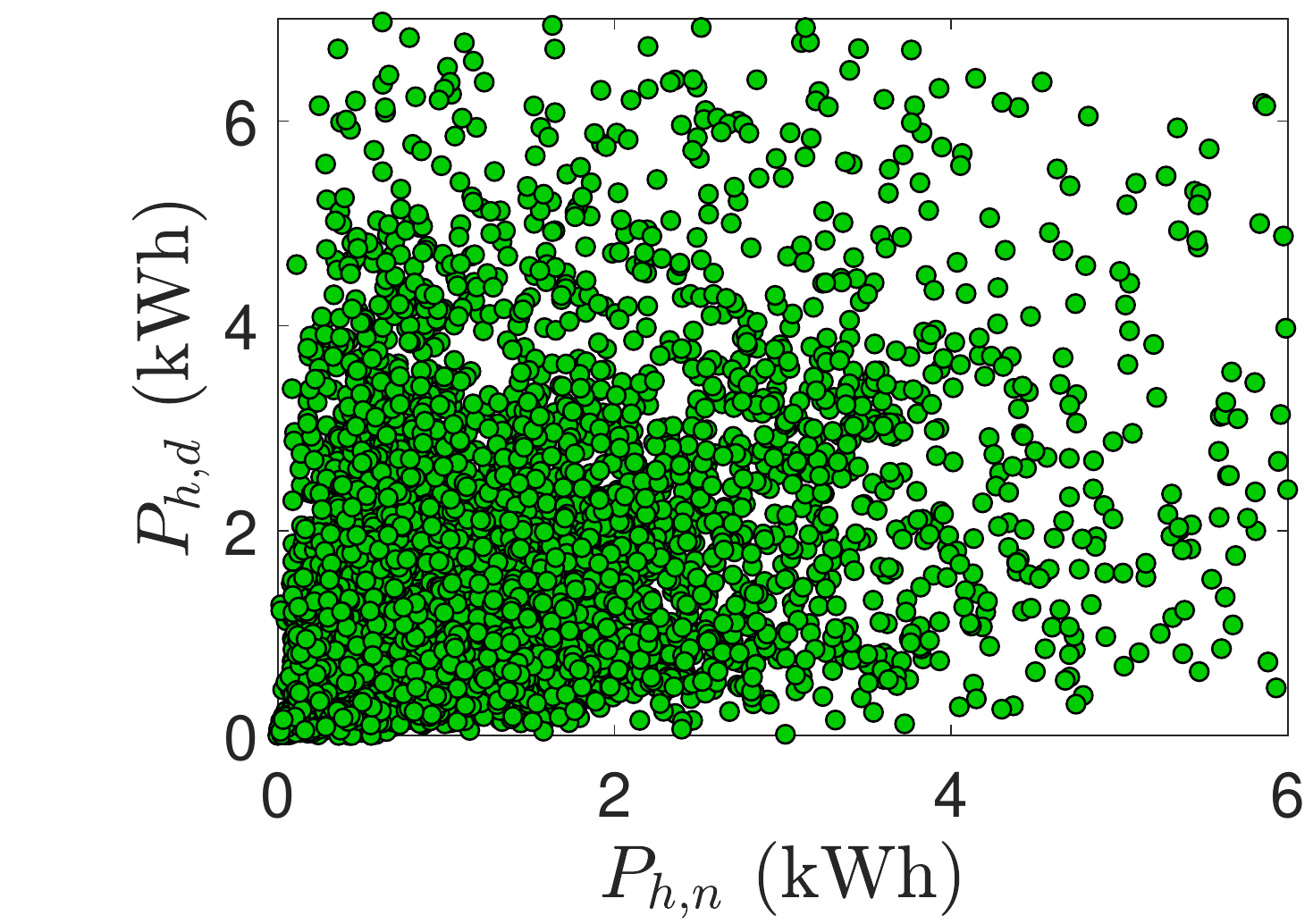}
}
\hfill
\subfloat[Customers w/o PV (Daily)\label{sfig:corr_demand_dy}]{
\includegraphics[width=0.41\linewidth]{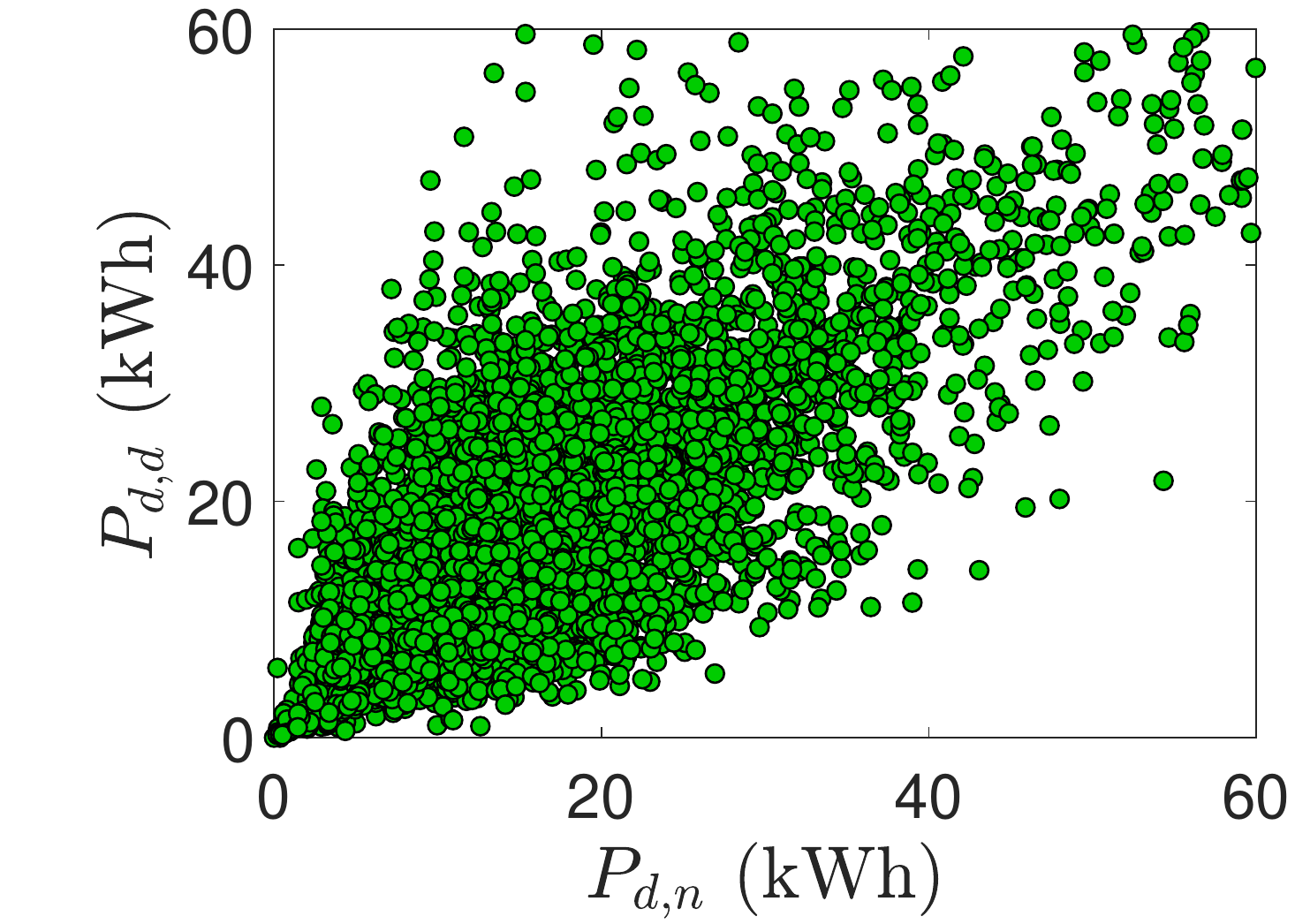}
}
\hfill
\subfloat[Customers w/o PV (Weekly)\label{sfig:corr_demand_wk}]{
\includegraphics[width=0.42\linewidth]{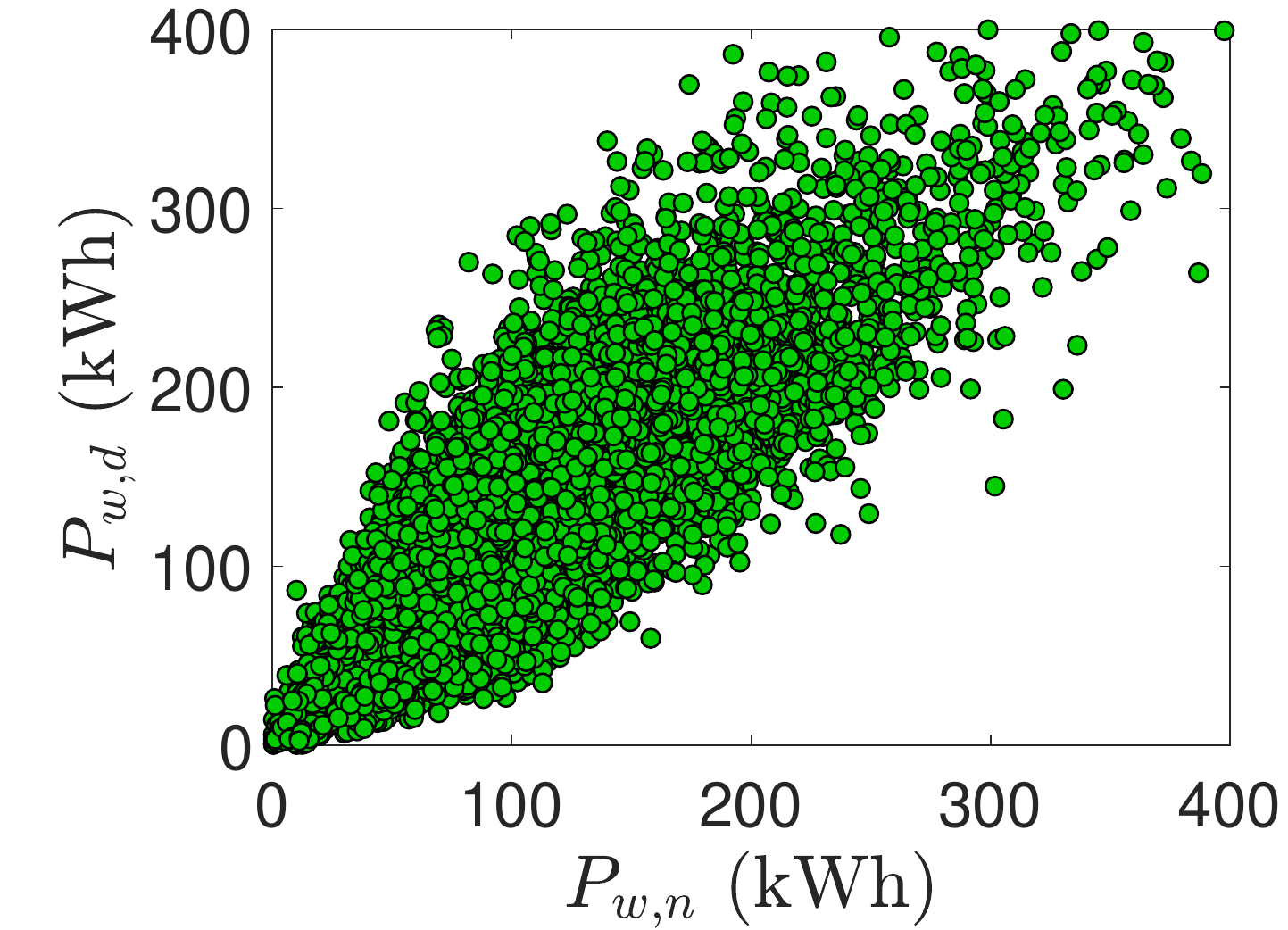}
}
\hfill
\subfloat[Customers w/o PV (Monthly)\label{sfig:corr_demand}]{
\includegraphics[width=0.4\linewidth]{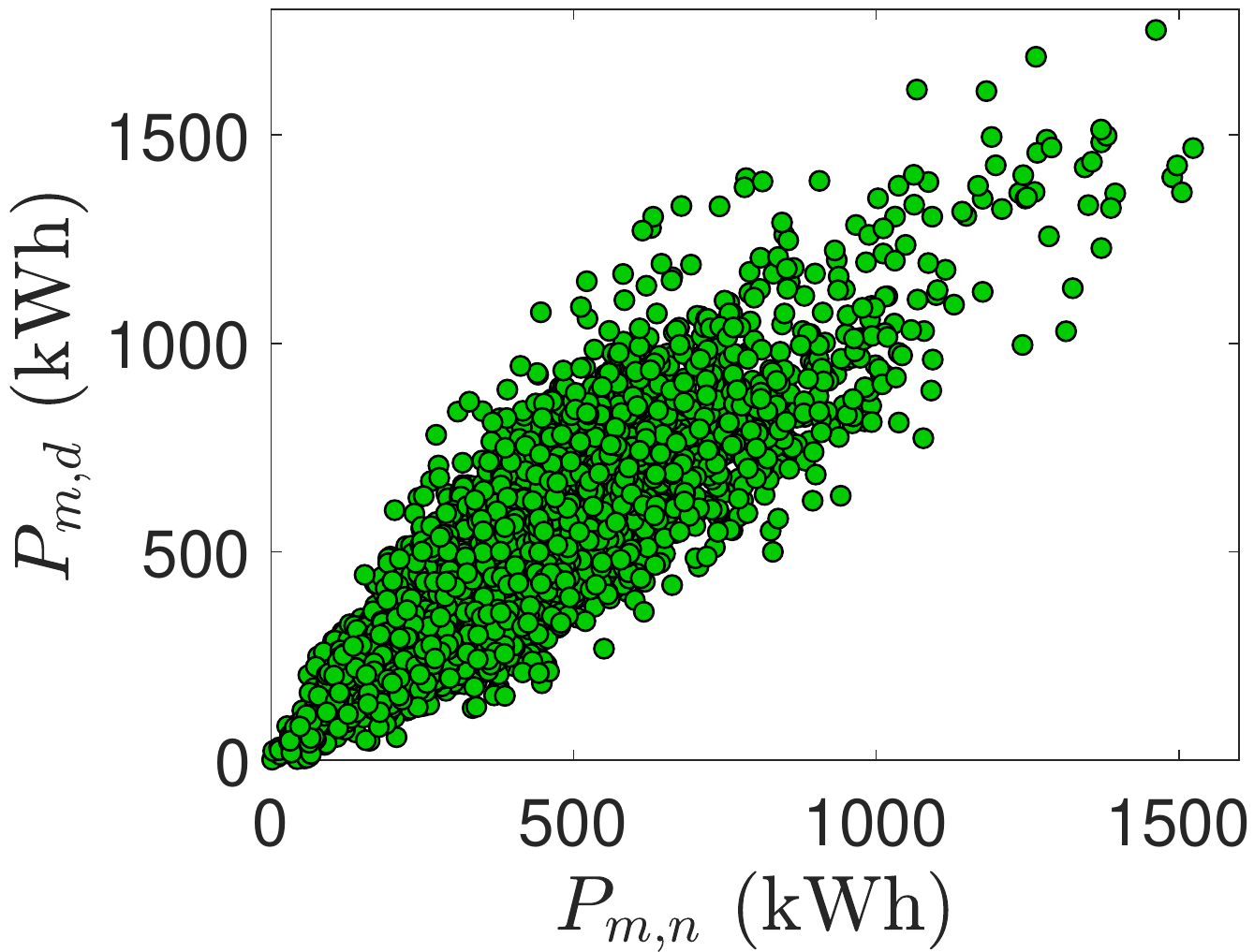}
}
\hfill
\subfloat[Customers w/ PV (Monthly)\label{sfig:corr_demand_W_solar}]{
\includegraphics[width=0.42\linewidth]{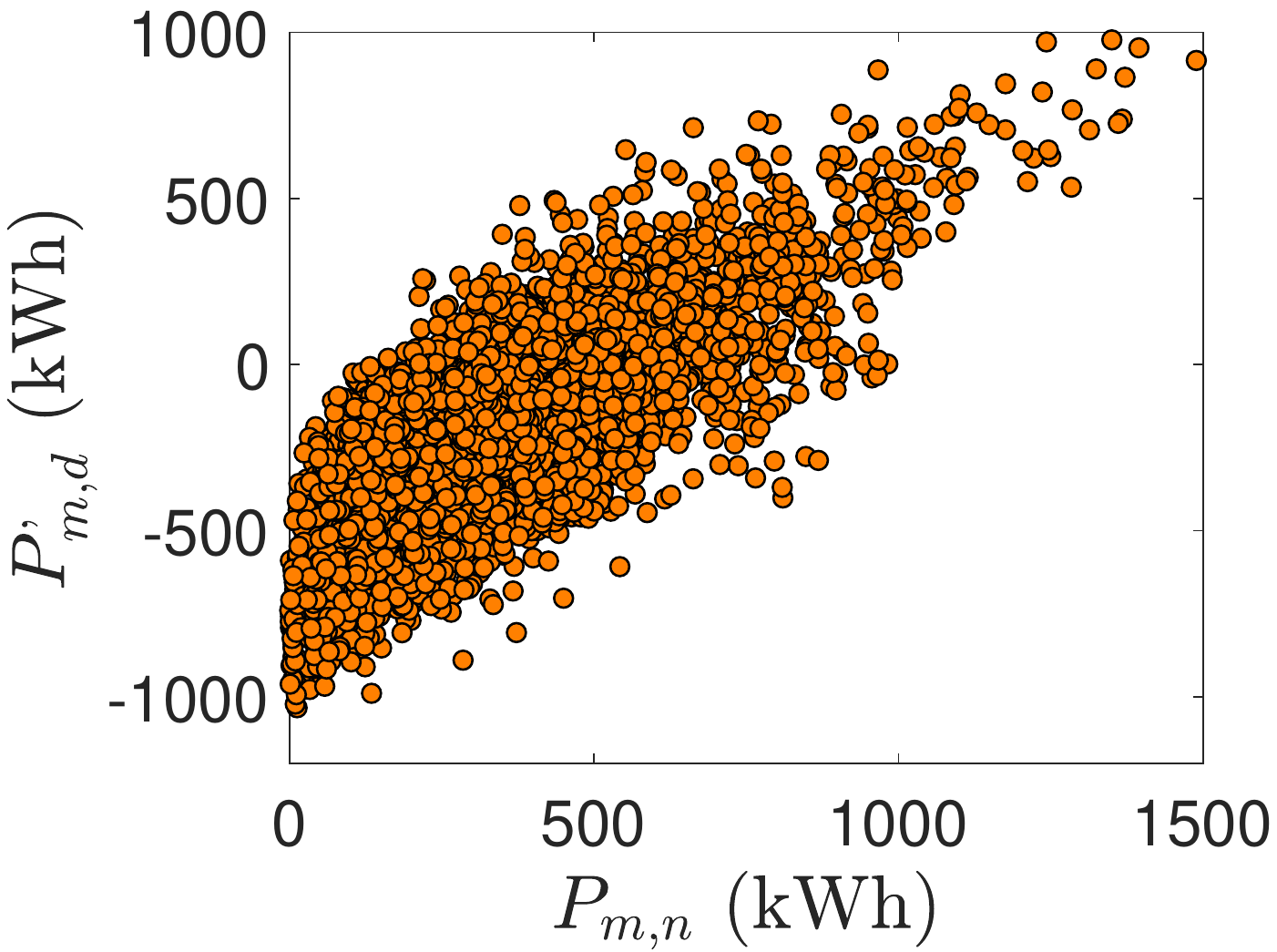}
}
\caption{Observations from real smart meter data.}
\label{fig:correlation}
\end{figure}

One key finding which sets the foundation for the proposed disaggregation approach is that the correlation between nocturnal native demand and the diurnal native demand increases as the observation timescale increases. This finding is illustrated in Fig. \ref{fig:correlation}, where, $P_{h,d}$, $P_{d,d}$, $P_{w,d}$, and $P_{m,d}$ denote the diurnal native demands measured on hourly, daily, weekly, and monthly basis, respectively. $P_{h,n}$, $P_{d,n}$, $P_{w,n}$, and $P_{m,n}$ denote the nocturnal native demands at the corresponding timescales, respectively. $P_{m,d}'$ denotes the monthly diurnal net demand of customers with PVs. Numerically, the correlation coefficients corresponding to Fig. \ref{sfig:corr_demand_hr}-\ref{sfig:corr_demand} are 0.56, 0.77, 0.89, and 0.91, respectively. In our paper, we employ the strong correlation of monthly native demand to perform disaggregation. The importance of this correlation is that it can be leveraged to reveal the monthly BTM generation of customers with PVs.  For instance, consider two customers, one with PV and one without PV. These two customers can have statistically-similar monthly nocturnal net demand, however, their monthly diurnal net demand will be significantly different from a statistical perspective due to BTM PV generation at daytime. Specifically, Fig. \ref{sfig:corr_demand_W_solar} shows the nocturnal-diurnal net demand distribution for customers with PV which is significantly different from Fig. \ref{sfig:corr_demand}. Thus, the distribution shown in Fig. \ref{sfig:corr_demand}, which represents the behavior of customers \textit{without} PV, can be used as a benchmark to determine whether a customer has BTM PV generation and estimate the monthly solar power. These findings have inspired us to construct a joint distribution of monthly nocturnal and diurnal \textit{native} demands of customers \textit{without} PVs to evaluate the deviation caused by the BTM PV generation of customers \textit{with} PVs. These deviations correspond to monthly BTM solar generation.

\subsection{Constructing the Nocturnal-Diurnal Native Demand PDF}\label{sec:GMM_detailed} 
We use a parametric PDF estimation technique known as GMM to construct the joint distribution of known monthly nocturnal and diurnal native demands of customers without PVs. A GMM is a linear combination of Gaussian components, and has demonstrated high flexibility and robustness in modeling arbitrary distributions \cite{GMM}. Since utilities have access to a large amount of native demand data, the constructed GMM-based joint PDF is able to probabilistically describes the quantitative relationship between the monthly nocturnal native demand and monthly diurnal native demand for customers without PVs. The native demand of customers \textit{with} PVs also follow this joint PDF, while their observed monthly net demand can deviate from the joint distribution. Compared with empirical histograms, the GMM-based PDF only has a limited number of parameters, therefore, it can be conveniently leveraged for estimating the BTM PV generation of the customers \textit{with} PVs. In our problem, the GMM approximation model can be described as follows:
\begin{equation}  \label{eq:GMM_1}
f(P_{m,n},P_{m,d}|\pmb{\Lambda})=\displaystyle\sum_{k=1}^{S}\theta_{k}g_k(P_{m,n},P_{m,d}|\pmb{\mu}_{k},\pmb{\Sigma}_{k}),
\end{equation}
where, $f(\cdot,\cdot)$ denotes the approximated joint PDF, $P_{m,n}$ and $P_{m,d}$ denote the monthly nocturnal and diurnal native demands of customers without PVs (i.e., customers belonging to $C_P$), respectively. $\pmb{\Lambda}$ denotes the parameter collection, $\{S, \theta_k, \pmb{\mu}_k, \pmb{\Sigma}_k\}$, which needs to be learned based on known native demand data. $S$ denotes the total number of Gaussian components. $\theta_k$'s are the weights corresponding to the bi-variate Gaussian components $g_k(\pmb{Z}|\pmb{\mu}_{k},\pmb{\Sigma}_{k})$ with $\pmb{Z}=[P_{m,n},P_{m,d}]$, which satisfy $\sum_{k=1}^{S}\theta_{k}=1$ and $0\leq \theta_k \leq 1$. The bi-variate Gaussian component is defined as
\begin{multline}  \label{eq:GMM_2}
g_k(\pmb{Z}|\pmb{\mu}_{k},\pmb{\Sigma}_{k})=\frac{1}{(2\pi)|\pmb{\Sigma}_{k}|^{1/2}}\\
\exp\Big\{-\frac{1}{2}(\pmb{Z}-\pmb{\mu}_{k})^\top\pmb{\Sigma}_{k}^{-1}(\pmb{Z}-\pmb{\mu}_{k})\Big\},
\end{multline}
where, $\pmb{\mu}_{k}$ and $\pmb{\Sigma}_{k}$ are the Gaussian component mean vector and covariance matrix, respectively.  

To learn $\pmb{\Lambda}$, first, a dataset is constructed based on smart meter measurements of customers in $C_P$. In practice, $P_{m,n}$ and $P_{m,d}$ of customers in $C_P$ are known to utilities and can be obtained from hourly smart meter readings in each month:
\begin{subequations}  \label{eq:aggregation_all}
\begin{equation}   \label{eq:aggregation_noct}
P_{m,n} = \displaystyle \sum_{t \in I_{n}}^{}P_h(t), 
\end{equation}    
\begin{equation}    \label{eq:aggregation_diur}
P_{m,d} = \displaystyle \sum_{t \in I_{d}}^{}P_h(t),
\end{equation}
\end{subequations}
where, $P_h(t)$ denotes the native demand reading at the $t$'th hour, $I_{n}$ and $I_{d}$ denote the sets of nighttime and daytime hours, respectively. Then, we can obtain the matrix of monthly demands by concatenating all customers' monthly native demand pairs:
\begin{equation}  \label{eq:ZZ}
\mathbf{Z} = [\mathbf{Z}(1),\cdots,\mathbf{Z}(N_c)]^\mathsf{T}
\end{equation}
where, $N_c$ denotes the total number of customers, and $\mathbf{Z}(j)$ denotes a matrix of monthly nocturnal and diurnal native demand pairs of the $j$'th customer which is organized as follows:
\begin{equation}  \label{eq:ZZZ}
\mathbf{Z}(j) = 
\left[
\begin{array}{ccc}
P_{m,n}(j,1) & P_{m,d}(j,1) \\
P_{m,n}(j,2) & P_{m,d}(j,2) \\
\vdots & \vdots \\
P_{m,n}(j,N_m) & P_{m,d}(j,N_m)
\end{array}
\right]^\mathsf{T} 
\end{equation}
where, $N_m$ is the total number of months. Then, we can obtain a dataset of observed monthly demand samples, $\{\pmb{Z}(1), \cdots, \pmb{Z}(N')\}$, through partitioning $\mathbf{Z}$ by rows, where, $N'=N_c \times N_m$.

Thus, the problem of GMM approximation boils down to finding optimal parameter collection $\pmb{\Lambda}^*$ that best fits the obtained dataset of monthly native demands, $\mathbf{Z}$, by assuming that the data samples are drawn independently from the underlying distribution. The most well-established idea for learning GMM parameters is to solve an optimization problem \cite{pattern_recog_book,ele_learning}, whereby the objective function can be formulated to maximize data likelihood, as follows:
\begin{equation}  \label{eq:GMM_3}
\max_{\pmb{\Lambda}} \quad
\displaystyle \prod_{i'=1}^{N'} f\big(\pmb{Z}(i')|\pmb{\Lambda}\big),
\end{equation}
By taking the logarithm of objective function, (\ref{eq:GMM_3}) is rewritten as follows:
\begin{equation}  \label{eq:GMM_4}
\max_{\pmb{\Lambda}} \quad
\displaystyle \sum_{i'=1}^{N'} \ln \Big\{f(\pmb{Z}(i')|\pmb{\Lambda}) \Big\}.
\end{equation}
The optimization problem in (\ref{eq:GMM_4}) is solved using the expectation-maximization algorithm \cite{pattern_recog_book}. 

Based on the identified optimal GMM parameter collection from (\ref{eq:GMM_4}), $\pmb{\Lambda}^*$, the joint PDF of monthly nocturnal and diurnal native demands can be specifically written as
\begin{equation}  \label{eq:GMM_5}
f(P_{m,n},P_{m,d})=\displaystyle\sum_{k=1}^{S^*}\theta_{k}^* g_k^*(P_{m,n},P_{m,d}),
\end{equation}
where,
\begin{multline}  \label{eq:cust_2}
g_k^*(P_{m,n}, P_{m,d})=\frac{1}{2\pi \sigma_{P_{m,n},k}^* \sigma_{P_{m,d},k}^* \sqrt{1-{\rho_{k}^*}^{2}}}\\ \exp \bigg\{-\frac{1}{2(1-{\rho_{k}^*}^2)}\Big[
\frac{(P_{m,n}-\mu_{P_{m,n},k}^*)^2}{{\sigma_{P_{m,n},k}^*}^2} +\frac{(P_{m,d}-\mu_{P_{m,d},k}^*)^2}{{\sigma_{P_{m,d},k}^*}^2} \\ - \frac{2\rho_{k}^*(P_{m,n}-\mu_{P_{m,n},k}^*)(P_{m,d}-\mu_{P_{m,d},k}^*)}{\sigma_{P_{m,n},k}^*\sigma_{P_{m,d},k}^*}   \Big] \bigg\},
\end{multline}
where, $S^*$ and $\theta_k^*$ are the learned number of mixture Gaussian components and mixture weights, respectively. $\mu_{P_{m,n},k}^*$, $\mu_{P_{m,d},k}^*$, $\sigma_{P_{m,n},k}^*$, $\sigma_{P_{m,d},k}^*$, and $\rho_k^*$ denote the learned mean, variance, and correlation of $P_{m,n}$ and $P_{m,d}$ for the $k$'th component, respectively.

Using GMM and the learned parameters, the joint distribution of monthly nocturnal and diurnal native demands is optimally represented. This joint distribution can serve as a benchmark for detecting and examining the discrepancy caused by BTM PV generation.

\section{Customer-Level Solar Disaggregation via MLE}\label{sec:MLE}  
\begin{figure}
\centering
\includegraphics[width=0.76\linewidth]{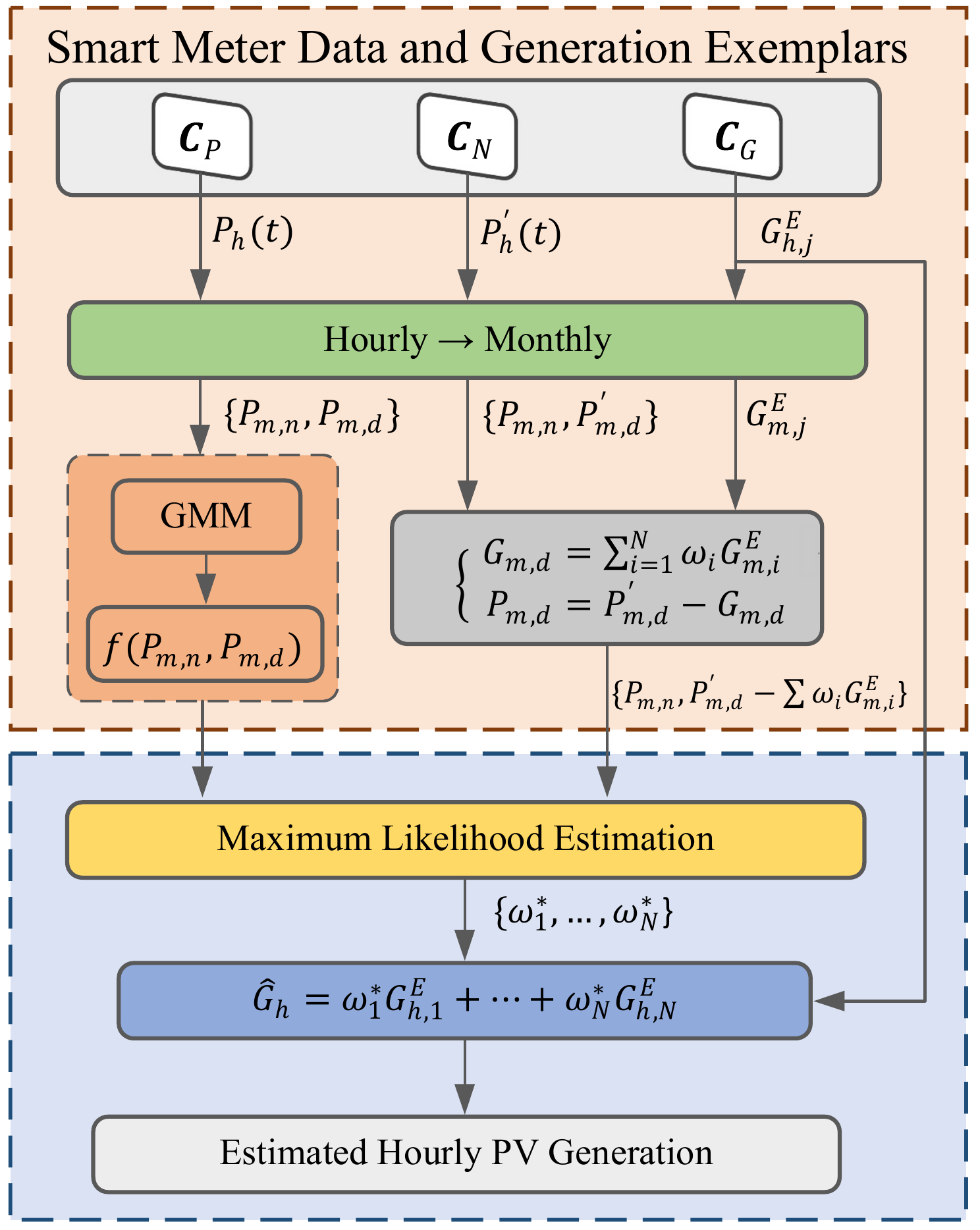}
\caption{Detailed structure of the proposed solar disaggregation approach for each customer.}
\label{fig:detaied_flow_chart}
\end{figure}

In this section, we disaggregate solar generation from net demand for \textit{each customer} with BTM PV using the constructed joint PDF, along with the measured net demand and solar exemplars. The detailed disaggregation process for each customer in $C_N$ is illustrated in Fig. \ref{fig:detaied_flow_chart}.

\subsection{MLE for Optimizing Solar Exemplar Weights}\label{sec:MLE_1}   
In a geographically bounded distribution system, it can be assumed that different PV arrays are subject to nearly identical meteorological inputs. Under this condition, the signature of an individual PV's generation profile is primarily determined by PV array's maximum power output and azimuth. The maximum power output determines the magnitude of generation curve \cite{kangping_li_solar}, and the azimuth determines the ``skewness" of generation profile \cite{indu_informatics}. Using the solar power curve of a south-facing PV array as a benchmark, the solar power curve of an east-facing PV array is left-skewed. A west-facing PV array has a right-skewed solar power curve. Therefore, the unknown BTM PV generation can be reliably represented using known generation profiles of BTM PVs (belonging to $C_G$) with typical orientations that serve as exemplars:
\begin{equation}  \label{eq:G_expression}
G_{m,d}=\sum_{i=1}^N \omega_i G_{m,i}^E = \pmb{\omega}^\mathsf{T} \pmb{G}_m^E,
\end{equation}
where, $N$ is the total number of solar exemplars, $\pmb{\omega}=[\omega_1, \cdots,\omega_N]^\mathsf{T}$ denotes an \textit{unknown} weight vector to be optimized, and $\pmb{G}_m^E=[G_{m,1}^E,\cdots,G_{m,N}^E]^\mathsf{T}$ denotes the PV generation vector of solar exemplars, where, $G_{m,i}^E$ is obtained by converting the known hourly diurnal PV generation into monthly diurnal solar power exemplars:
\begin{equation}    \label{eq:aggregation_PV_exemplar}
G_{m,i}^E=\sum_{t \in I_{d}}^{}G_{h,i}^{E}(t),
\end{equation}
where, $G_{h,i}^{E}(t)$ is the PV generation of the $i$'th exemplar at the $t$'th hour. Therefore, disaggregating BTM PV generation of each customer in $C_N$ comes down to finding optimal coefficients assigned to known solar exemplars. To do this, first, we represent the unknown monthly diurnal native demand using the known monthly net demand and monthly PV generation of solar exemplars:
\begin{equation}  \label{eq:Y_CN}
P_{m,d}=P_{m,d}'-\pmb{\omega}^\mathsf{T}\pmb{G}_m^E.
\end{equation}
where, $P_{m,d}'$ is the known monthly net demand which can be obtained as follows:
\begin{equation}    \label{eq:aggregation_net_P}
P_{m,d}'=\sum_{t \in I_{d}}^{}P_h'(t),
\end{equation}
where, $P_h'(t)$ denotes the recorded net demand at the $t$'th hour.

\begin{algorithm}[t]
\caption{Disaggregating BTM PV generation and native demand from net demand for \textit{each customer}}\label{alg:algorithm_proposed}
\begin{algorithmic}[1]
\State {Classify residential customers into three types: $C_P$, $C_G$, and $C_N$}
\Procedure{Data Conversion}{}
    \State {For customers in $C_P$:}
    \State {$P_{m,n} \gets \sum_{t \in I_n}^{} P_h(t)$, $P_{m,d} \gets \sum_{t \in I_d}^{} P_h(t)$}
    \State {For customers in $C_G$:}
    \State {$G_{m,i}^E \gets \sum_{t \in I_d}^{} G_{h,i}^{E}(t) \quad i=1,\cdots,N$} 
    \State {For customers in $C_N$:}
    \State {$P_{m,n} \gets \sum_{t \in I_n}^{} P_h'(t)$, $P_{m,d}' \gets \sum_{t \in I_d}^{} P_h'(t)$}
\EndProcedure
\Procedure{Construct Nocturnal-Diurnal Native Demand PDF}{}
    \State {For customers in $C_P$:}
    \State {$\pmb{\Lambda} \gets \{\theta_k, \pmb{\mu}_k, \pmb{\Sigma}_k\} \quad k=1,\cdots,S$}
    \State {$\pmb{\Lambda}^* \gets \underset{\pmb{\Lambda}}{\textit{max}} \sum_{i'=1}^{N'} \ln \{f(P_{m,n},P_{m,d}|\pmb{\Lambda}) \}$}
\EndProcedure
\Procedure{Perform MLE for Optimizing Weights}{}
    \State {For customers in $C_N$:}
    \State {$P_{m,d} \gets P_{m,d}'- \pmb{\omega}^\mathsf{T}(\pmb{G}_m^E)$}
    \State {Solve optimization in (\ref{eq:overall}) to obtain $\pmb{\omega}^*$}
\EndProcedure
\Procedure{Estimate Hourly BTM PV Generation and Native Demand}{}
    \State {For customers in $C_N$:}
    \State {$\pmb{\hat{G}}_h \gets (\pmb{\omega}^*)^\mathsf{T} {\mathbf{G}_h^E}, \pmb{\hat{P}}_h \gets \pmb{P}_h' - \pmb{\hat{G}}_h$}
\EndProcedure
\end{algorithmic}
\end{algorithm}

Since the monthly nocturnal and diurnal native demands of customers \textit{with} PVs probabilistically follow the constructed GMM-based joint PDF, by substituting (\ref{eq:Y_CN}) into (\ref{eq:GMM_5}), we can represent the distribution function for customers with BTM PVs as follows:
\begin{equation}    \label{eq:optimization_1}
f\big(P_{m,n}, P_{m,d}'-\pmb{\omega}^\mathsf{T}\pmb{G}_m^E \big).
\end{equation}
Note that (\ref{eq:G_expression})-(\ref{eq:optimization_1}) apply to each month, and we do not add the dimension of month into these equations for the sake of conciseness.
Then, the exemplar weight optimization is formulated as an MLE problem over all months, as described as follows:
\begin{equation}    \label{eq:MLE_0}
\pmb{\omega}^* = \max_{\pmb{\omega}} \Big\{\prod_{t'=1}^{M} f(P_{m,n}(t'), P_{m,d}'(t'), \pmb{G}_m^E(t')| \pmb{\omega}) \Big\},
\end{equation}
where, $M$ is the total number of months. 

Further, the optimization solution should be subject to multiple constraints to enforce the identified PV generation to be non-positive and the estimated native demand to be non-negative. Finally, by taking logarithm of (\ref{eq:MLE_0}) and introducing the constraints, the complete optimization problem is elaborated as follows:
\begin{subequations}  \label{eq:overall}
\begin{flalign}  \label{eq:overall_a}
\underset{\pmb{\omega}}{\textit{max}} \;
\Big\{\sum_{t'=1}^{M} \ln \big[  f(P_{m,n}(t'), P_{m,d}'(t'), \pmb{G}_m^E(t')| \pmb{\omega})  \big]\Big\} - \frac{1}{2} \lambda ||\pmb{\beta}||_2^2,  && 
\end{flalign}
\vspace{-12pt} 
\begin{flalign}  \label{eq:overall_b}
    \; \textit{s.t.} \quad  \;\;
    (\pmb{\omega}^\mathsf{T}  \mathbf{G}_h^E)^\mathsf{T} \leq  \pmb{0},  &&
\end{flalign}
\vspace{-20pt}  
\begin{flalign}  \label{eq:overall_c}
\ \quad\quad\quad  
\pmb{P}_h' - (\pmb{\omega}^\mathsf{T}  \mathbf{G}_h^E)^\mathsf{T} \ge \pmb{\beta},   &&
\end{flalign}
\vspace{-20pt}
\begin{flalign}  \label{eq:overall_d}
\ \quad\quad\quad  
\pmb{\beta} \le  \pmb{0},   &&
\end{flalign}
\end{subequations}
where, $\mathbf{G}_h^E=[\pmb{G}_h^E(1),\cdots,\pmb{G}_h^E(N_h)]$ denotes a matrix of hourly PV generation solar exemplars' time series, $\pmb{G}_h^E(\tau)=[G_{h,1}^E(\tau),\cdots,G_{h,N}^E(\tau)]^\mathsf{T},\tau=1,\cdots,N_h$ denotes the vector of solar exemplars' generation readings at the $\tau$'th hour, $N_h$ denotes the total number of hourly demand readings, $\pmb{P}_h'$ denotes the time-series hourly net demand readings and $\pmb{0}$ represents a zero vector. In addition to maximizing the likelihood function shown in (\ref{eq:MLE_0}), a $l_2$-norm penalty term, $- \frac{1}{2} \lambda ||\pmb{\beta}||_2^2$, is added into the objective function, where, $\lambda \ge0$ is a tuning parameter and $\pmb{\beta}$ is a vector with non-positive elements. Constraint (\ref{eq:overall_b}) ensures that the estimated hourly PV generation is non-positive. Constraints (\ref{eq:overall_c}) and (\ref{eq:overall_d}) ensure that the estimated time-series native demand is larger than a non-positive vector whose $l_2$-norm is penalized in the objective function. This penalty term is based on the following consideration: In practice, it is common for the solar generation to have data quality problems. For example, PV arrays can stop running due to solar panel failures, and the recorded anomalous samples are usually smaller than the unrecorded expected values. For the customers whose PV generation is supposed to be disaggregated from the known net demand, the unwanted PV failure does not cause significant disaggregation error. This is because the relatively smaller anomalous PV generation samples cause an unwanted rise in the net demand readings only for a limited number of samples. These larger net demand readings can still give us positive estimated native demand values, since the native demand is estimated by subtracting the disaggregated BTM PV generation from net demand. In comparison, the anomalous readings of \textit{solar exemplars} can cause a negative estimated native demand, which brings significant estimation errors. This is because removing a zero or near-zero PV generation from a negative net demand measurement gives us a negative estimated native demand value. Thus strictly constraining the estimated native demand to be non-negative can cause unwanted errors. Therefore, we have leveraged a soft margin to penalize the effect of anomalous data. Since the purpose of introducing the penalty term is to allow for a small number of negative native demand estimates, the value of tuning parameter, $\lambda$, should be chosen in a way to ensure that the number of negative native demand estimates is close to the number of solar exemplars' anomalous data samples. The MLE problem in (\ref{eq:overall}) is solved via numerical optimization using interior-point methods.

\begin{figure}[htbp]
\centering
\subfloat[Empirical histogram\label{sfig:j1}]{
\includegraphics[width=0.75\linewidth]{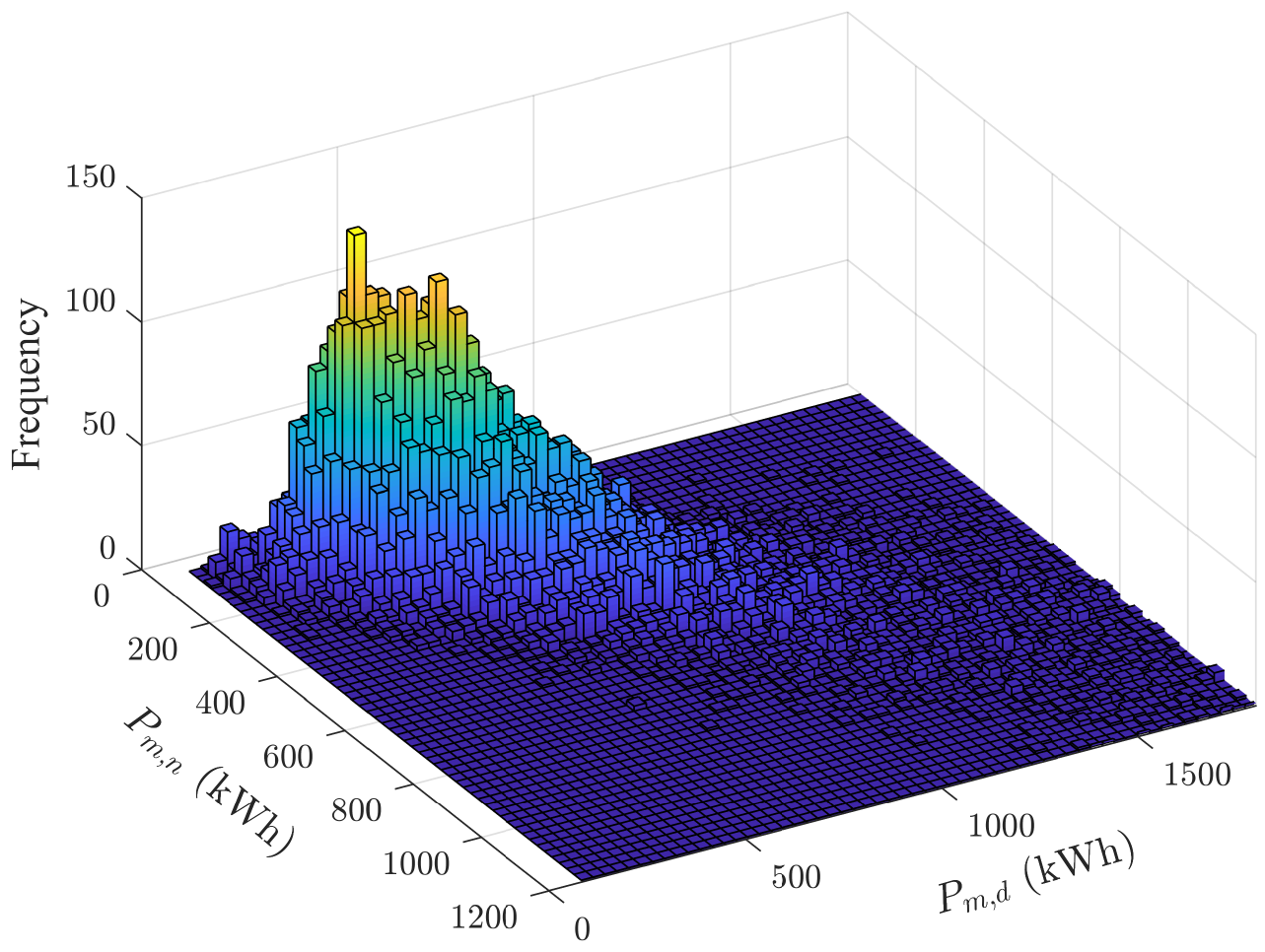}
}
\hfill
\subfloat[GMM-based estimation\label{sfig:j2}]{
\includegraphics[width=0.75\linewidth]{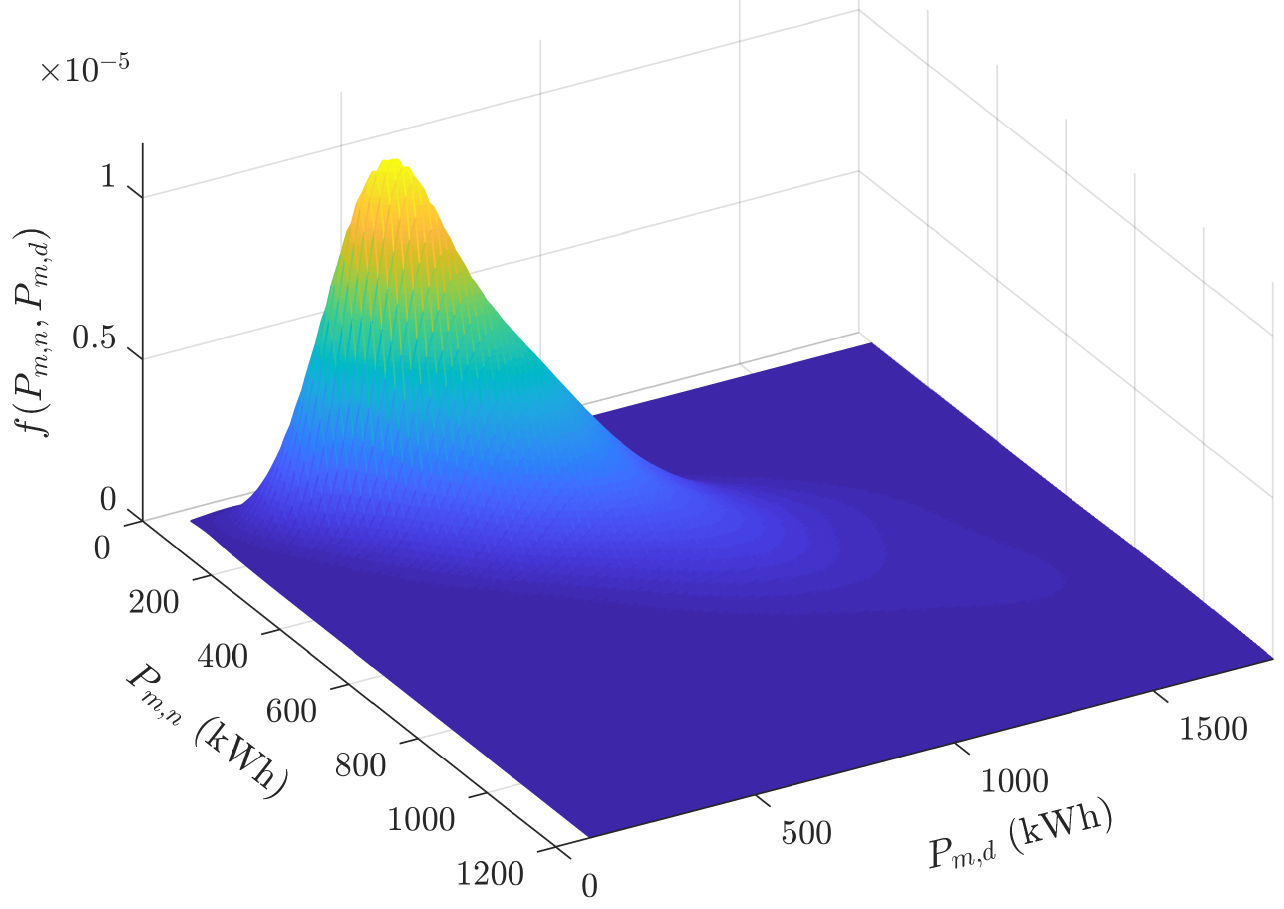}
}
\caption{Joint PDF estimation of monthly nocturnal and diurnal native demands.}
\label{fig:joint}
\end{figure}

\begin{figure*}[htbp]
\centering
\subfloat[PV generation\label{sfig:solar_curves}]{
\includegraphics[width=0.9\linewidth]{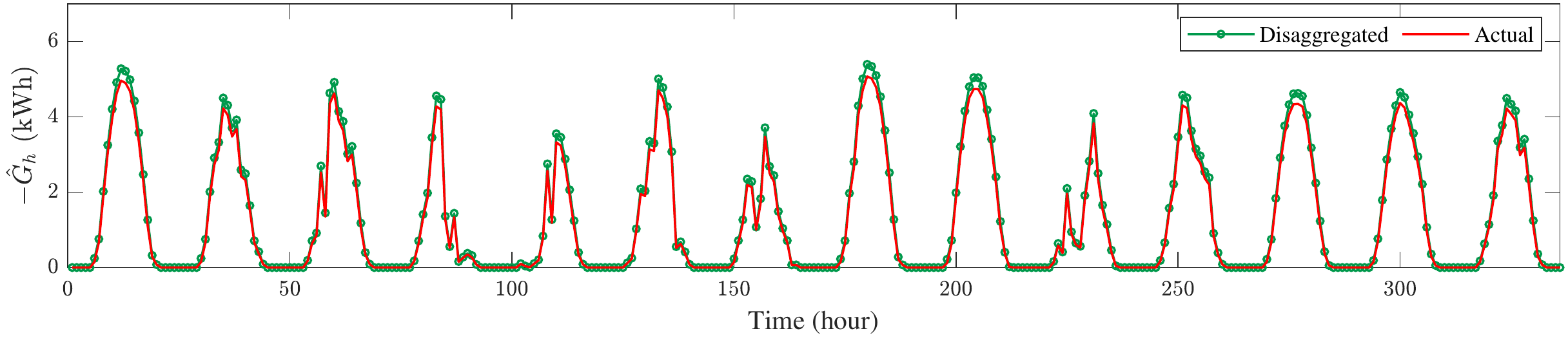}
}
\hfill
\subfloat[Native demand\label{sfig:demand_curves}]{
\includegraphics[width=0.9\linewidth]{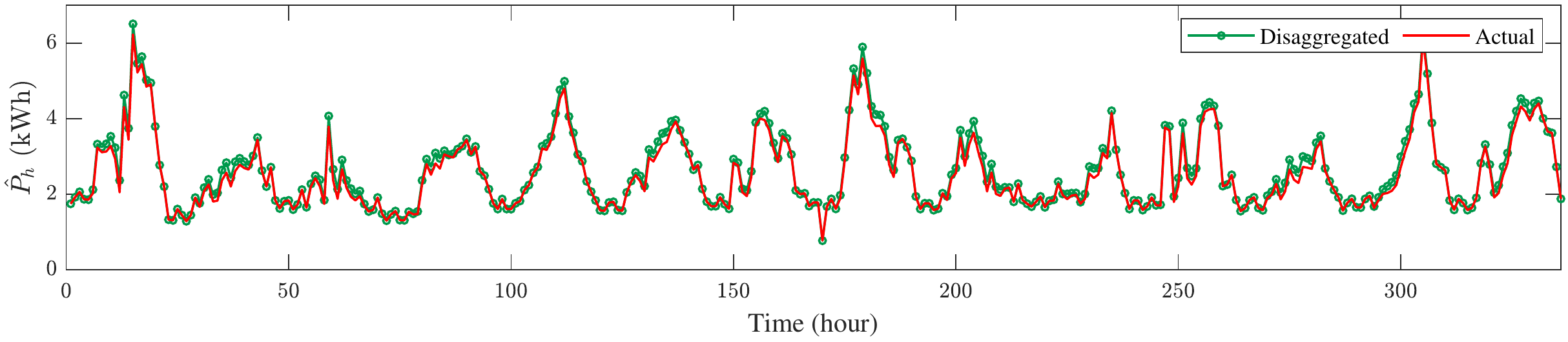}
}
\caption{Two-week disaggregated PV generation and native demand curves, along with corresponding actual curves.}
\label{fig:P_G_curves}
\end{figure*}

\subsection{Estimating Hourly PV Generation and Native Demand}\label{sec:MLE_2} 
By solving the optimization (\ref{eq:overall}), we can obtain the optimized weight vector, $\pmb{\omega}^*$, which is utilized to estimate the unknown hourly BTM PV generation of customers with PVs:
\begin{equation}  \label{eq:estimated_G}
\pmb{\hat{G}}_h=(\pmb{\omega}^*)^\mathsf{T}\mathbf{G}_h^E.
\end{equation}
Further, the hourly native demand can be estimated by subtracting the disaggregated BTM PV generation from known net demand readings:
\begin{equation}  \label{eq:estimated_P}
\pmb{\hat{P}}_h = \pmb{P}_h' - \pmb{\hat{G}}_h.
\end{equation}

An algorithmic overview of the aforementioned steps of BTM PV generation disaggregation is summarized in Algorithm \ref{alg:algorithm_proposed}.

\section{Case Study}\label{sec:casestudy}
In this section, the proposed customer-level rooftop BTM solar power separation approach is verified using real smart meter and PV generation data described in Section \ref{sec:overall}. 


\subsection{Assessing Statistical Behavior of Customers}
The empirical histogram and the GMM-based estimation of $f(P_{m,n},P_{m,d})$ are shown in Fig. \ref{sfig:j1} and Fig. \ref{sfig:j2}, respectively. Comparing these two figures, it can be seen that GMM is able to accurately model the joint distribution of monthly nocturnal and diurnal native demands using smooth parametric Gaussian density functions. Also note that the joint PDF surface is quite narrow, i.e., the data is highly concentrated around the linear representative of nocturnal and diurnal demands. This corroborates the high correlation between monthly nocturnal and diurnal native demands observed in Fig. \ref{sfig:corr_demand}.

\subsection{BTM PV Generation Disaggregation Validation}\label{sec:casestudy_2}
Using the constructed GMM-based joint PDF, along with the known monthly net demand of customers with PVs and PV generation of solar exemplars, we can solve the MLE problem described in (\ref{eq:overall}). When selecting solar exemplars, it is demonstrated that on average, three exemplars can sufficiently represent the PV generation profiles, and introducing additional solar exemplars does not bring further disaggregation accuracy improvement \cite{Fankun_Bu}. Thus, we have selected three typical solar power curves from $C_G$ corresponding to PVs facing east, south and west, respectively. Fig. \ref{fig:P_G_curves} shows disaggregated PV generation and native demand curves of one customer over two weeks, along with corresponding actual profiles. In Fig. \ref{sfig:solar_curves}, it can be seen that the disaggregated curve closely fits the actual profile, regardless of the solar volatility on some days. This shows the accurate diaggregation capability of our proposed method and also corroborates our observation that PV generation profiles with similar PV array orientations are highly correlated. Fig. \ref{sfig:demand_curves} shows the disaggregated and actual native demand profiles. It can be observed that despite the uncertain and volatile pattern of native demand, the disaggregated curve can still fit the real profile. 

\begin{figure}[htbp]
\centering
\subfloat[Hourly\label{sfig:cluster_hourly}]{
\includegraphics[width=0.42\linewidth]{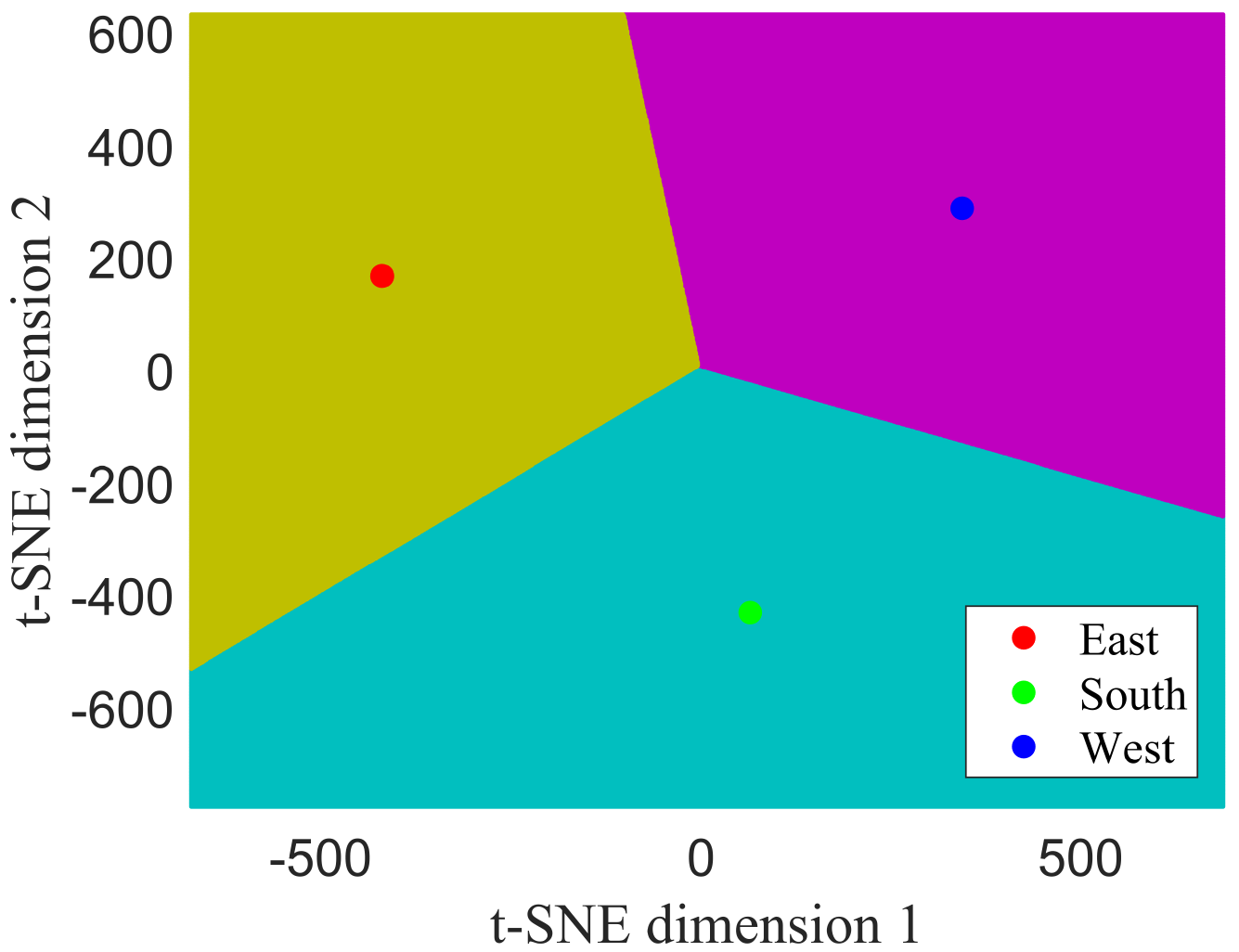}
}
\hfill
\subfloat[Monthly\label{sfig:cluster_monthly}]{
\includegraphics[width=0.42\linewidth]{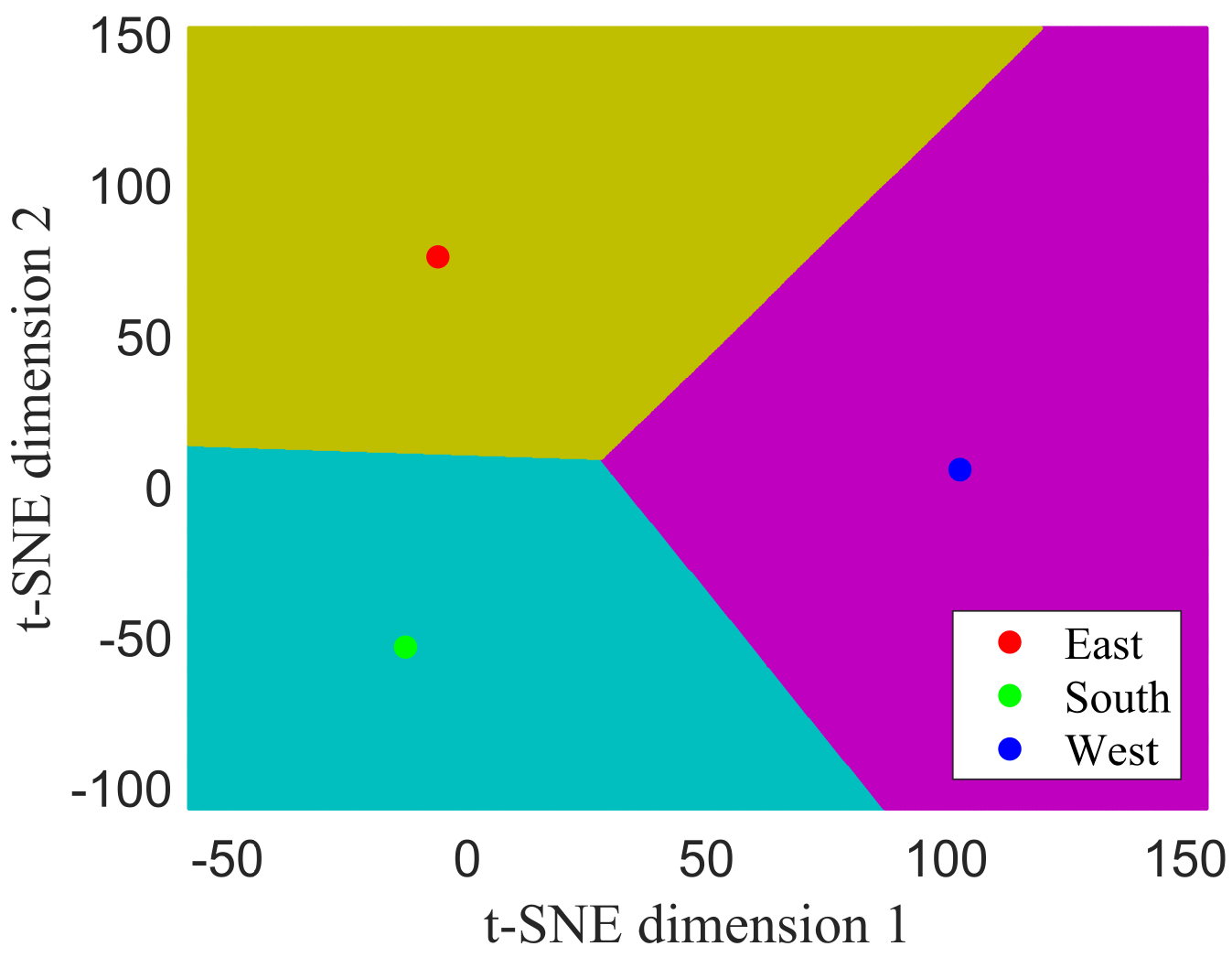}
}
\caption{Visualizing the distinguishability of time-series PV generation curves of solar exemplars.}
\label{fig:cluster}
\end{figure}

\begin{figure}[htbp]
\centering
\subfloat[Solar exemplars\label{sfig:exemplars}]{
\includegraphics[width=0.86\linewidth]{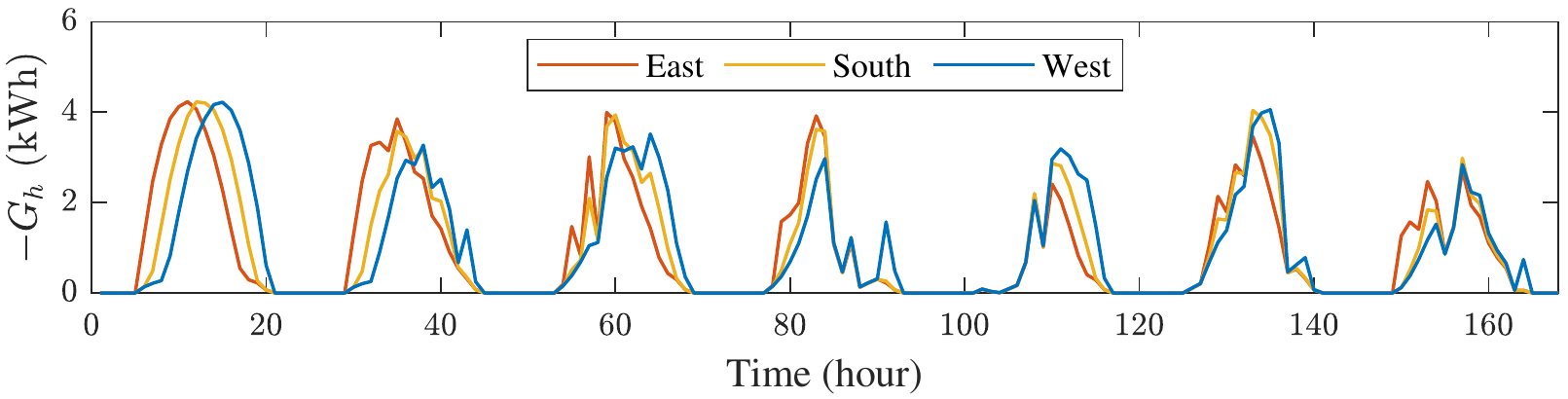}
}
\hfill
\subfloat[A PV facing east\label{sfig:east}]{
\includegraphics[width=0.9\linewidth]{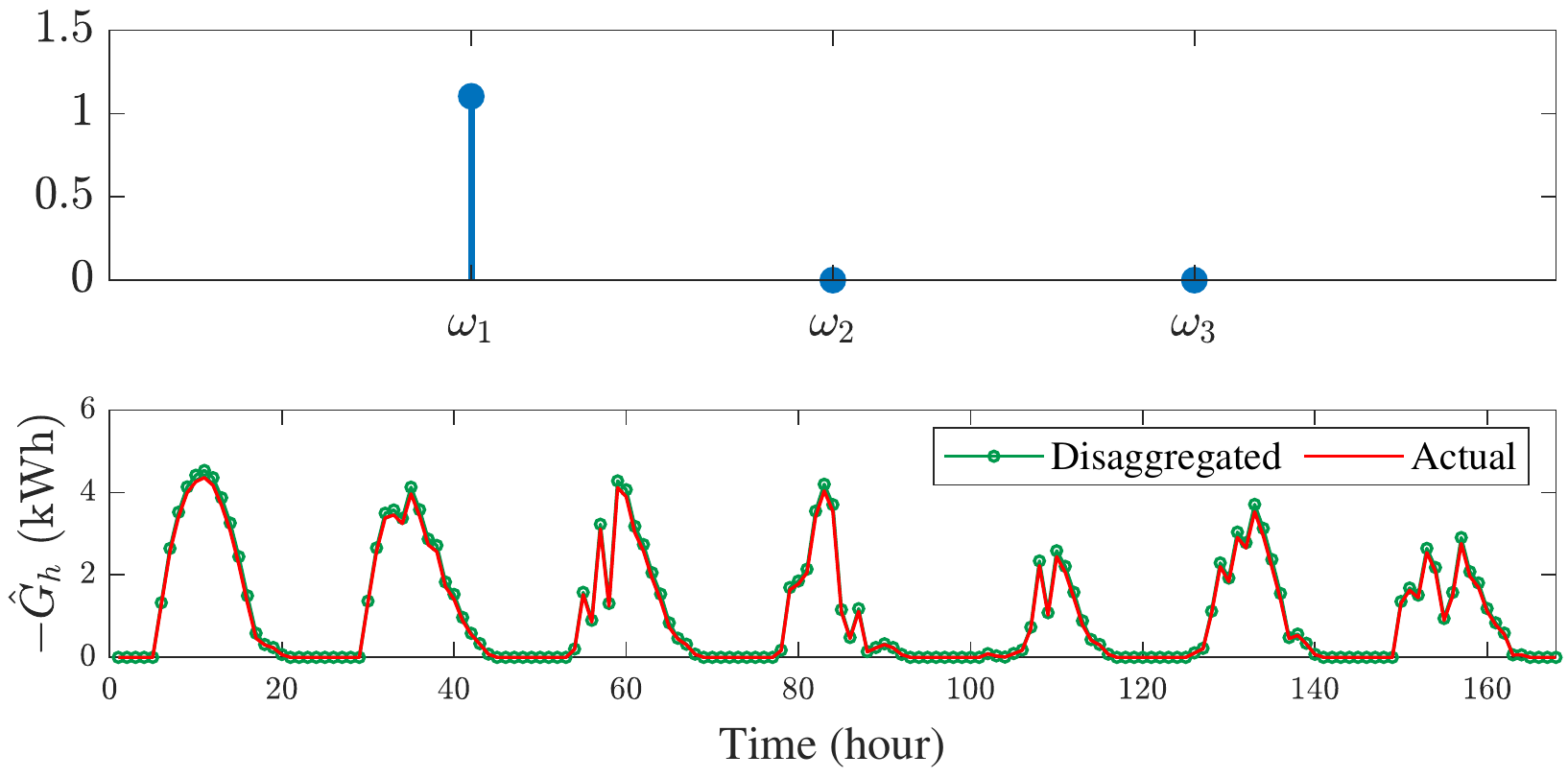}
}
\hfill
\subfloat[A PV facing south\label{sfig:south}]{
\includegraphics[width=0.9\linewidth]{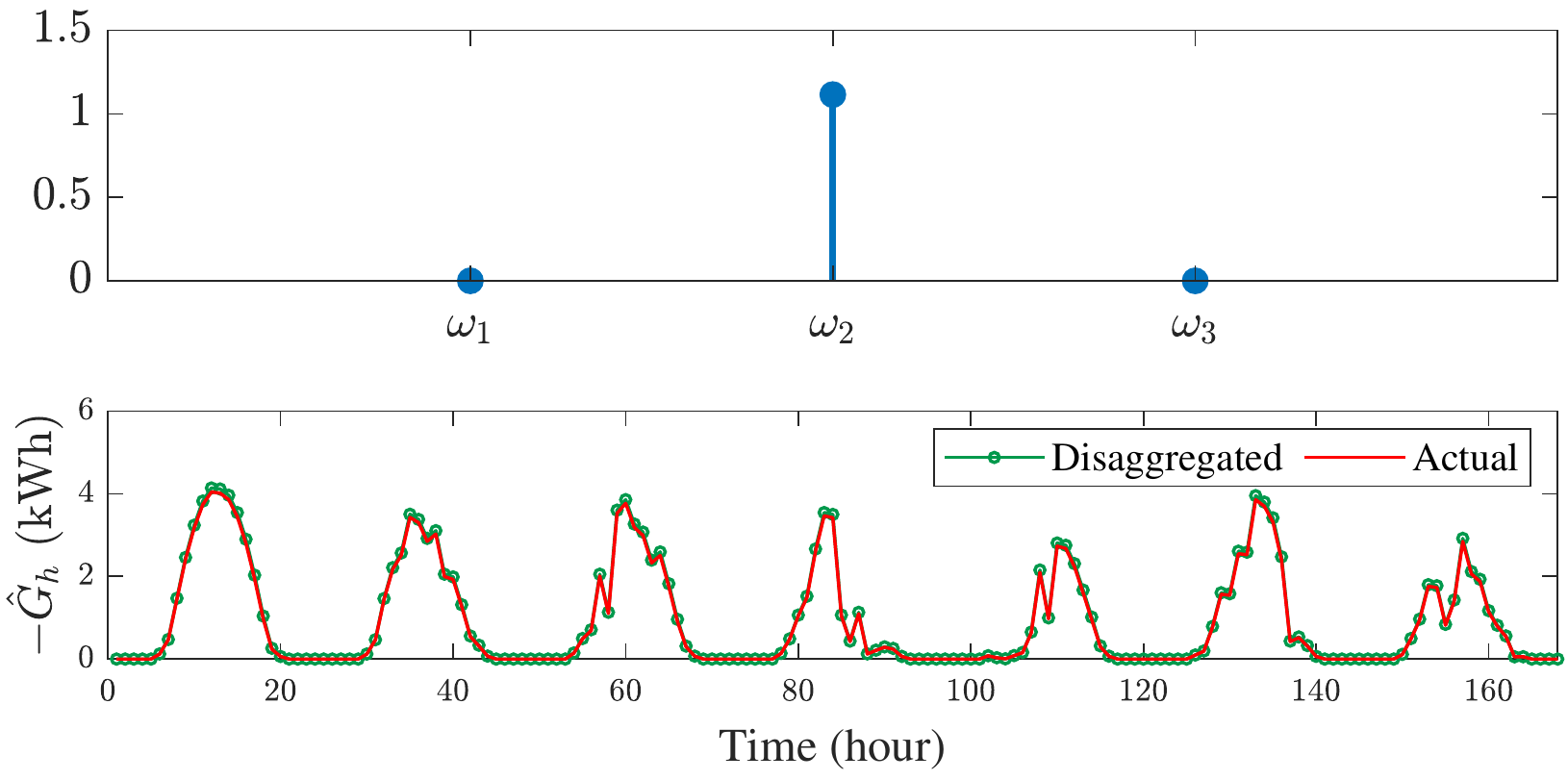}
}
\hfill
\subfloat[A PV facing west\label{sfig:west}]{
\includegraphics[width=0.9\linewidth]{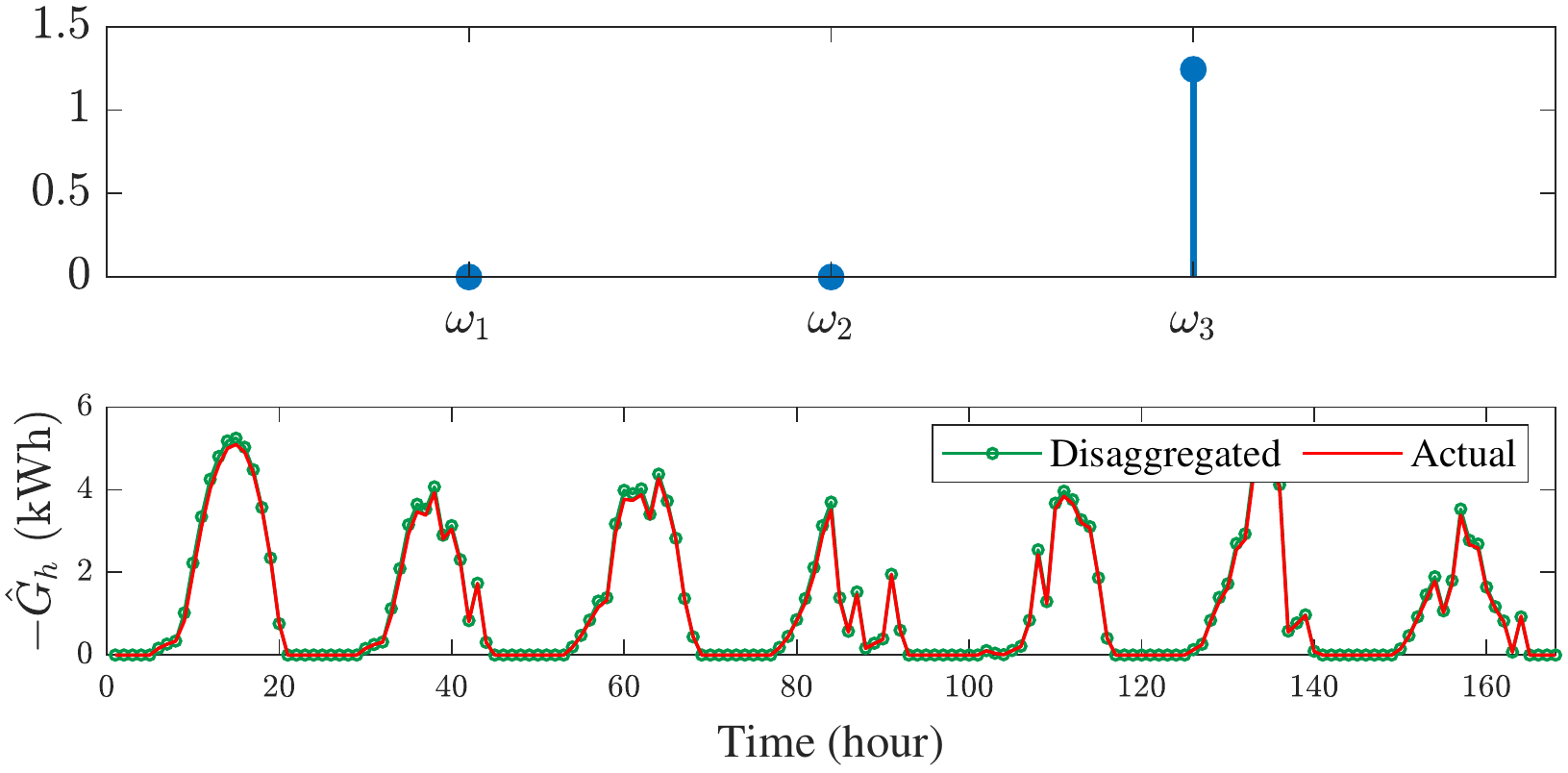}
}
\caption{The proposed approach can correctly track proper solar exemplars to perform disaggregation.}
\label{fig:weights_comparison}
\end{figure}

It is of importance to examine the representative feature of typical solar exemplars. In (\ref{eq:G_expression}), the unknown BTM PV generation is represented using known generation profiles of solar exemplars. Therefore, these PV generation profiles which serve as exemplars should be distinguishable, otherwise, multiple solutions of weights with the same losses can be derived in the MLE optimization process. We have employed a dimensionality reduction technique known as t-SNE to visualize the dissimilarities among PV generation profiles of solar exemplars \cite{t_SNE}. Note that each time point is treated as one dimension in our problem. The dimensions of hourly and monthly PV generation time series are reduced for convenient visualization, as shown in Fig. \ref{fig:cluster}. Fig. \ref{sfig:cluster_hourly} shows the reduced two-dimensional solar power exemplars based on the hourly PV generation of PVs facing east, south and west. As can be seen, the solar exemplars are demonstrated to be distinct. Similarly, the monthly PV generation of solar exemplars also demonstrate distinguishable features, as shown in Fig. \ref{sfig:cluster_monthly}. This is consistent with our observation that solar generation profiles are primarily determined by PV panel orientations in geographically bounded distribution systems.

It is of significance to test whether the proposed approach can track the appropriate exemplars (east, south or west) in the disaggregation process. Fig. \ref{sfig:exemplars} shows PV generation curves of the three exemplars facing east, south and west. We can see that PVs with different orientations show distinct profile skewness. Fig. \ref{sfig:east} shows the disaggregated and real PV generation curves of a PV facing east, along with the optimized weights assigned to the three solar exemplars. It can be seen that the weight corresponding to the first exemplar (i.e., PV facing east) is much larger compared to the other two weights, which validates the tracking ability of our proposed approach. This verification can also be observed in Fig. \ref{sfig:south} and \ref{sfig:west}, which show the weights, disaggregated and actual PV generation curves of PVs facing south and west, respectively. In all cases, our method has accurately detected the correct underlying BTM PV panel orientations.

\begin{figure}[htbp]
\centering
\subfloat[PV generation\label{sfig:MAPE_1}]{
\includegraphics[width=0.46\linewidth]{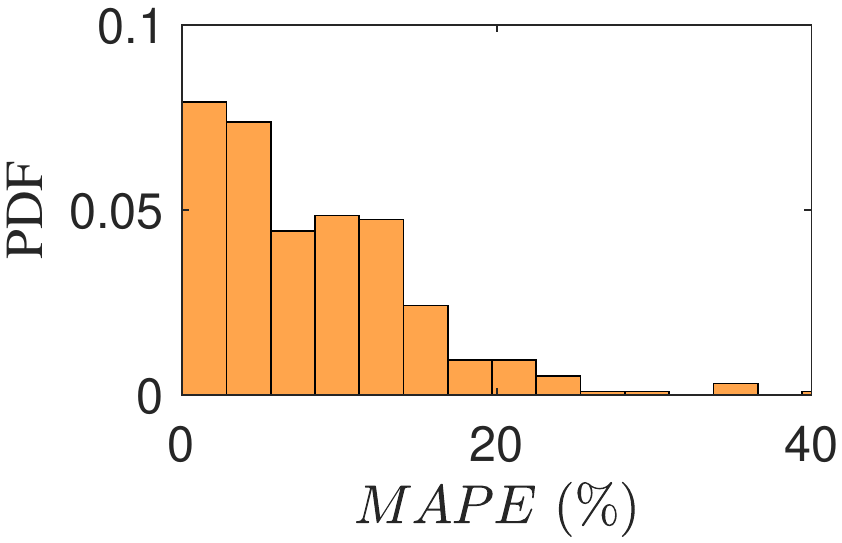}
}
\hfill
\subfloat[Native demand\label{sfig:MAPE_2}]{
\includegraphics[width=0.46\linewidth]{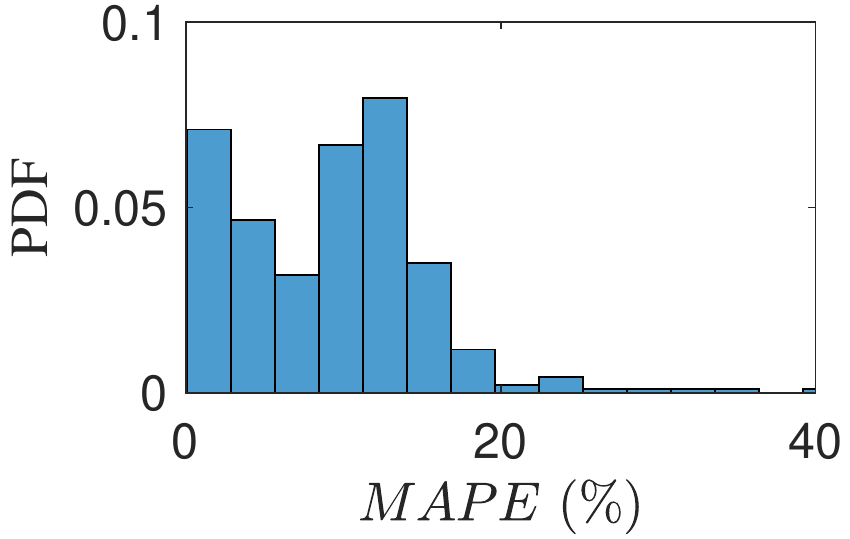}
}
\caption{Empirical distribution of $MAPE$ of disaggregated  estimates.}
\label{fig:MAPE_distribution}
\end{figure}

\begin{figure}[htbp]
\centering
\subfloat[PV generation\label{sfig:RMSE_1}]{
\includegraphics[width=0.43\linewidth]{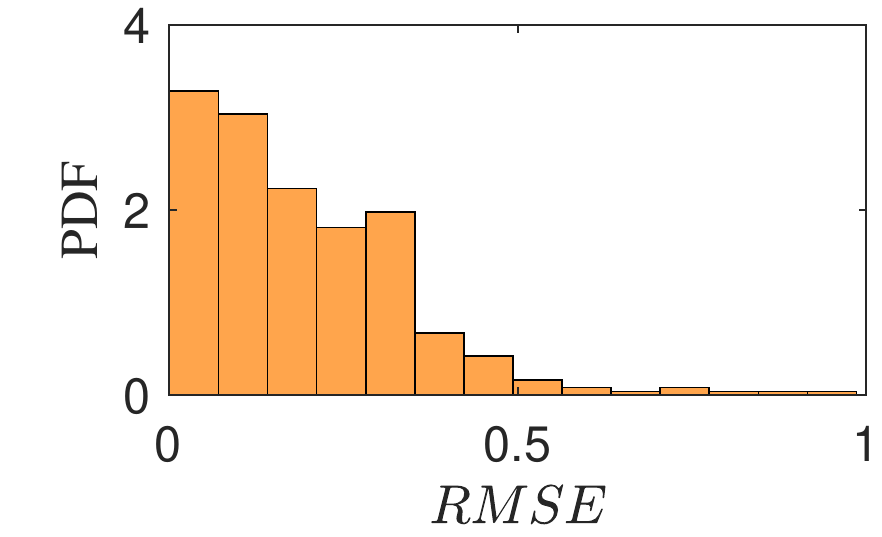}
}
\hfill
\subfloat[Native demand\label{sfig:RMSE_2}]{
\includegraphics[width=0.42\linewidth]{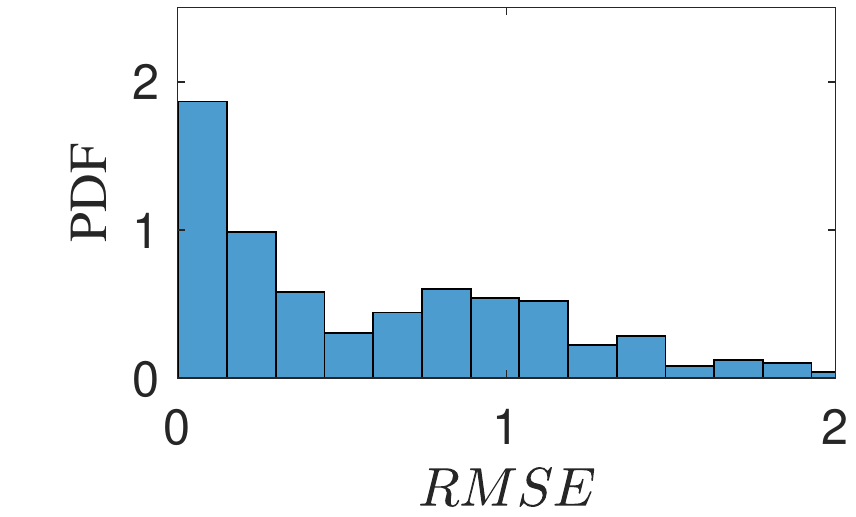}
}
\caption{Empirical distribution of $RMSE$ of disaggregated estimates.}
\label{fig:RMSE_distribution}
\end{figure}

The proposed customer-level BTM solar separation approach is applied to all 237 customers with PVs, and the disaggregation accuracy for each customer is evaluated in terms of mean absolute percentage error ($MAPE$), which is calculated as follows:
\begin{equation}   \label{eq:MASE_eq} 
MAPE=\frac{100\%}{N_h'}\cdot \sum_{t=1}^{N_h'} \Bigg| \frac{\hat{O}_h(t)-{O}_h(t)}{\frac{1}{N_h'} \sum_{t=1}^{N_h'}|{O}_h(t)|}  \Bigg|
\end{equation}
where, $N_h'$ denotes the total number of non-zero PV generation observations for an individual customer, $O_h$ can be $P_h$ or $G_h$. Fig. \ref{fig:MAPE_distribution} shows the distribution of disaggregation error for all customers in terms of $MAPE$. As can be seen, majority of the $MAPE$s are less than 20\%. This effectively demonstrates the generalization ability of our proposed method. Table \ref{tbl:MAPE} summarises the empirical cumulative distribution function (CDF) of disaggregation $MAPE$. As can be seen, for the disaggregated hourly PV generation, 80\% of the $MAPE$s are less than 13.5\%. Regarding the disaggregated hourly native demand, 80\% of the $MAPE$s are less than 14.9\%. This effectively verifies the disaggregation accuracy of our proposed approach.

The disaggregation accuracy for each customer is also evaluated using $RMSE$, which is computed as follows:
\begin{equation}   \label{eq:RMSE_eq} 
RMSE=\sqrt{\frac{\Sigma_{t=1}^{N_h^{'}}(\hat{O}_h(t)-{O}_h(t))^2}{N_h^{'}}}.
\end{equation}
Fig. \ref{fig:RMSE_distribution} shows the empirical distributions of the $RMSE$ of disaggregated estimates based on all customers' computed $RMSE$s. It can be seen that most PV generation and native demand $RMSE$s are less than 0.5 and 1.5, respectively. Also, the empirical CDF of disaggregation $RMSE$ is calculated for a comprehensive examination, as shown in Table \ref{tbl:RMSE}.

\begin{table}[htbp]
\centering
	\renewcommand{\arraystretch}{1.5}
	\setlength{\tabcolsep}{4.8pt}
	\caption{Empirical CDF of Disaggregation $MAPE$}\label{tbl:MAPE}
	\begin{tabular}{cccccc}
	    \toprule[1pt]
		Empirical CDF & 0.2 & 0.4 & 0.6 & 0.8 & 1.0  \\
	    \hline
        $MAPE$ of $\hat{G}_{h}$ (\%) & 2.5 & 4.8 & 9.7 & 13.5 & 33.4 \\
        $MAPE$ of $\hat{P}_{h}$ (\%) & 3.1  & 8.3 & 12.3  & 14.9  & 29.1\\
        \bottomrule[1pt]
	\end{tabular}
\end{table}

\begin{table}[htbp]
\centering
	\renewcommand{\arraystretch}{1.5}
	\setlength{\tabcolsep}{4.8pt}
	\caption{Empirical CDF of Disaggregation $RMSE$}\label{tbl:RMSE}
	\begin{tabular}{cccccc}
	    \toprule[1pt]
		Empirical CDF & 0.2 & 0.4 & 0.6 & 0.8 & 1.0  \\
	    \hline
        $RMSE$ of $\hat{G}_{h}$ & 0.06 & 0.12 & 0.21 & 0.31 & 4.51 \\
        $RMSE$ of $\hat{P}_{h}$ & 0.10  & 0.28 & 0.73  & 1.08  & 3.85\\
        \bottomrule[1pt]
	\end{tabular}
\end{table}

\subsection{Testing the Robustness of the Proposed Approach}
\begin{figure}
\centering
\includegraphics[width=0.75\linewidth]{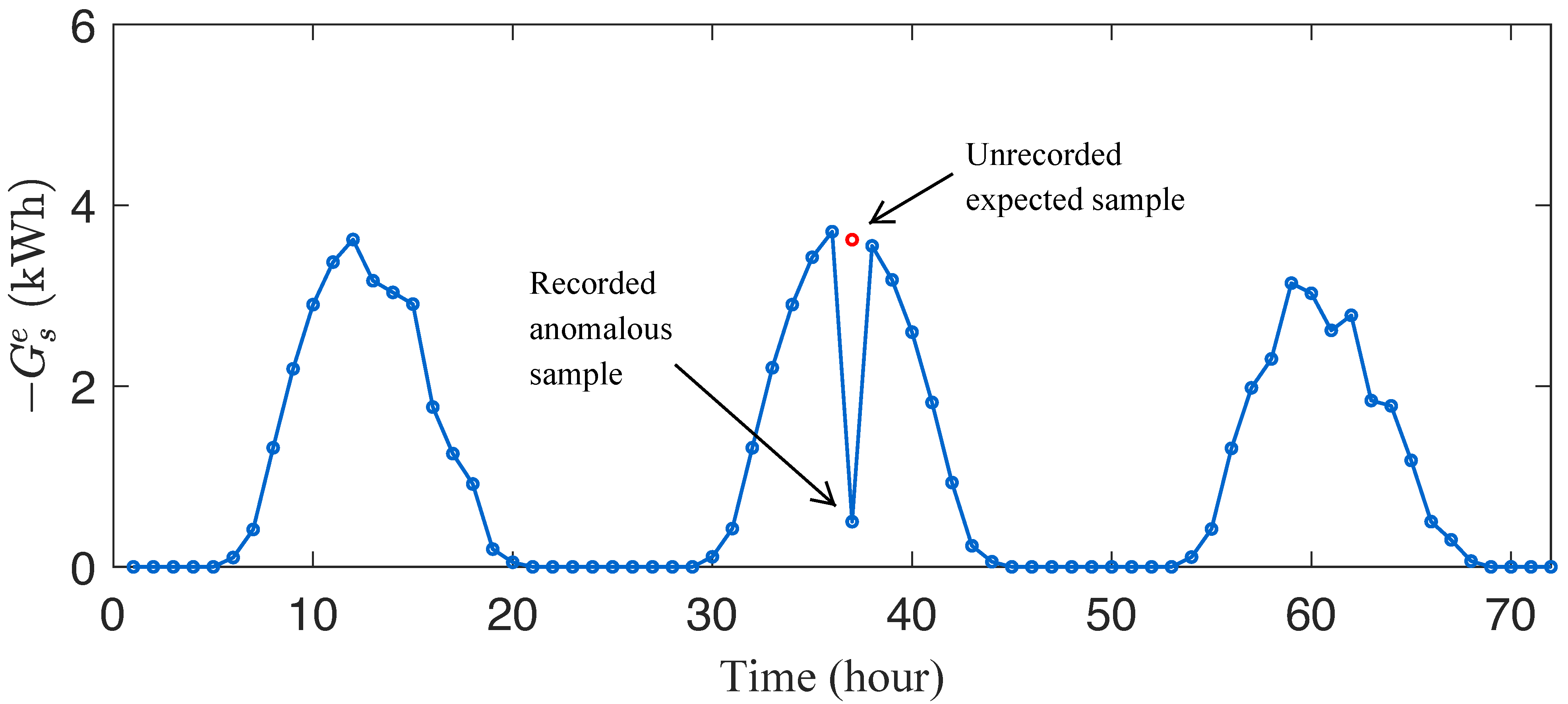}
\caption{A solar exemplar with an anomalous sample due to PV failure.}
\label{fig:dad_data}
\end{figure}

\begin{figure}[htbp]
\centering
\subfloat[PV generation\label{sfig:disa_PV_comp}]{
\includegraphics[width=0.8\linewidth]{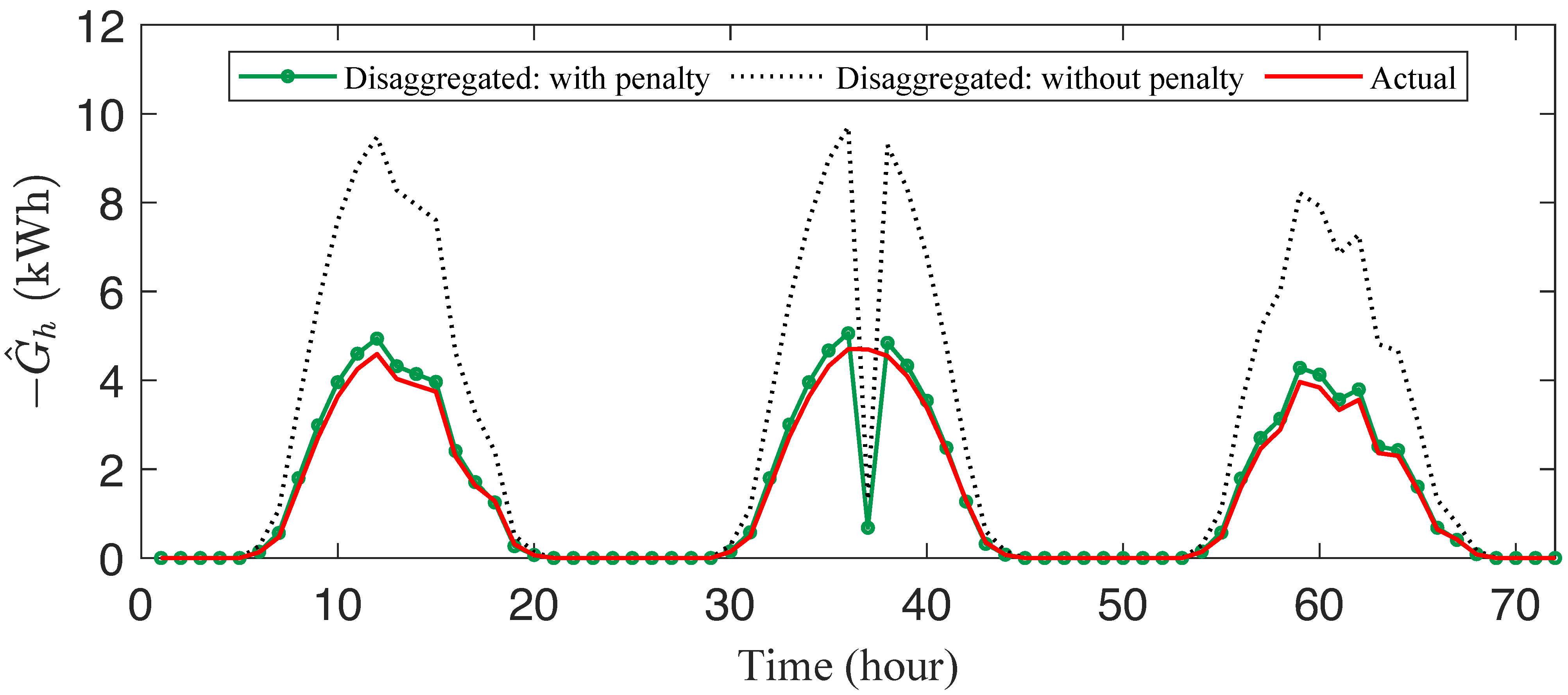}
}
\hfill
\subfloat[Native demand\label{sfig:disa_demand_comp}]{
\includegraphics[width=0.8\linewidth]{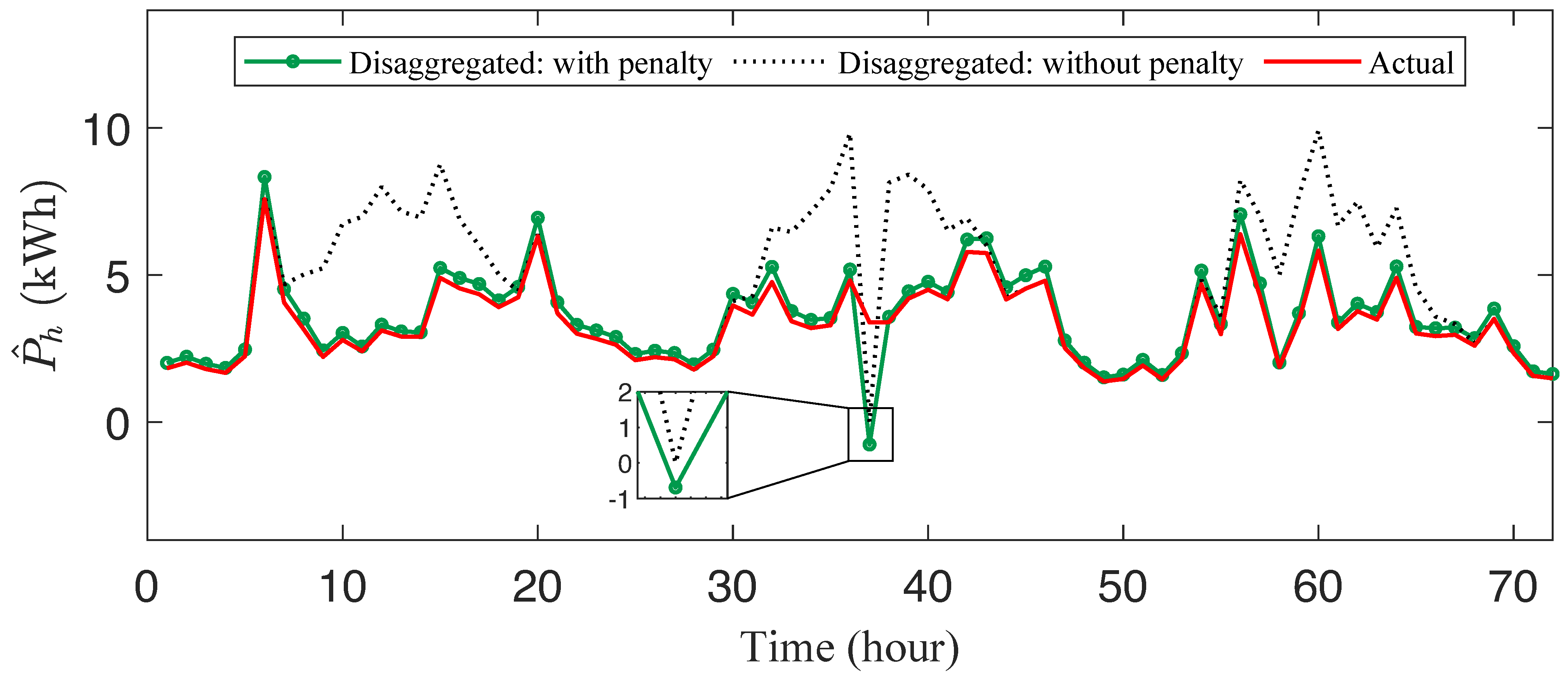}
}
\caption{The introduction of penalty term significantly improves disaggregation accuracy and robustness.}
\label{fig:P_G_disaggregation_comparison}
\end{figure}

It is common for a practical metering system to have a small number of anomalous measurements in solar exemplars, as shown in Fig. \ref{fig:dad_data}, where the unrecorded expected generation is denoted as a red circle. The typical reason for anomalous solar power data samples is PV failure, which causes the recorded data samples to be smaller than the unrecorded expected values. As previously elaborated in Section \ref{sec:MLE}, a penalty term is included in (\ref{eq:overall}) to mitigate the effect of solar exemplar's anomalous samples. Therefore, it is crucial to test the usefulness of the penalization mechanism. Note that the results in Section \ref{sec:casestudy_2} are obtained using (\ref{eq:overall}) with a penalty term. Thus, to conduct a performance comparison, we alter (\ref{eq:overall}) to obtain a new optimization formulation with the penalty term omitted, as expressed as follows:
\begin{subequations}  \label{eq:overall_new}
\begin{flalign}  \label{eq:overall_new_a}
\underset{\pmb{\omega}}{\textit{max}} \quad
\sum_{t'=1}^{M} \ln \big[  f(P_{m,n}(t'), P_{m,d}'(t'), \pmb{G}_m^E(t')| \pmb{\omega})  \big],  && 
\end{flalign}
\vspace{-12pt} 
\begin{flalign}  \label{eq:overall_new_b}
    \; \textit{s.t.} \quad  \;\;
    (\pmb{\omega}^\mathsf{T}  \mathbf{G}_h^E)^\mathsf{T} \leq  \pmb{0},  &&
\end{flalign}
\vspace{-20pt}  
\begin{flalign}  \label{eq:overall_new_c}
 \quad\quad\quad  
\pmb{P}_h' - (\pmb{\omega}^\mathsf{T}  \mathbf{G}_h^E)^\mathsf{T} \ge \pmb{0}.   &&
\end{flalign}
\end{subequations}
Then, using the solar exemplar with an anomalous sample in Fig. \ref{fig:dad_data}, we  utilize (\ref{eq:overall}) and (\ref{eq:overall_new}) to perform disaggregation, respectively.
Fig. \ref{fig:P_G_disaggregation_comparison} compares three-day disaggregated PV generation and native demand curves based on (\ref{eq:overall}) and (\ref{eq:overall_new}), respectively. The actual solar power and native demand curves are also plotted as benchmarks. In Fig. \ref{sfig:disa_PV_comp}, it can be seen that the disaggregated PV generation curve using (\ref{eq:overall}) can closely fit the actual curve except for at the hour that the solar exemplar's anomalous sample appears. In comparison, the disaggregated PV generation curve using (\ref{eq:overall_new}) significantly deviates from the actual benchmark. Regarding the disaggregated native demand, we can draw the same conclusion by observing Fig. \ref{sfig:disa_demand_comp}. The overestimation of PV generation and native demand using (\ref{eq:overall_new}) is due to the constraint that forces the estimated native demand to be strictly non-negative, as shown in Fig. \ref{sfig:disa_demand_comp}. In contrast, our approach presented in (\ref{eq:overall}) allows a negative native demand estimate to mitigate the anomalous samples' impact. To sum up, the introduction of penalty into the MLE optimization significantly enhances the robustness of our proposed approach against anomalous data.

\subsection{Performance Comparison}
\begin{figure}[htbp]
\centering
\subfloat[PV generation\label{sfig:MAPE_1_NPY}]{
\includegraphics[width=0.4\linewidth]{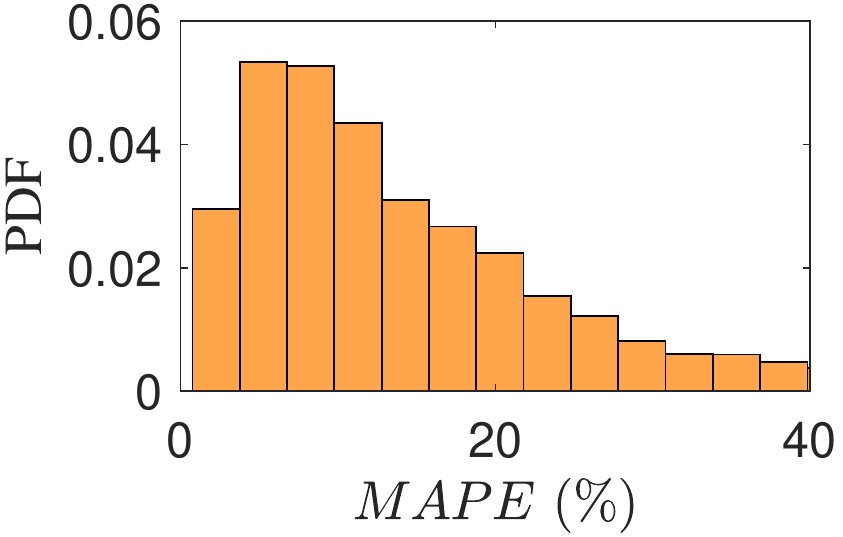}
}
\hfill
\subfloat[Native demand\label{sfig:MAPE_2_NPY}]{
\includegraphics[width=0.4\linewidth]{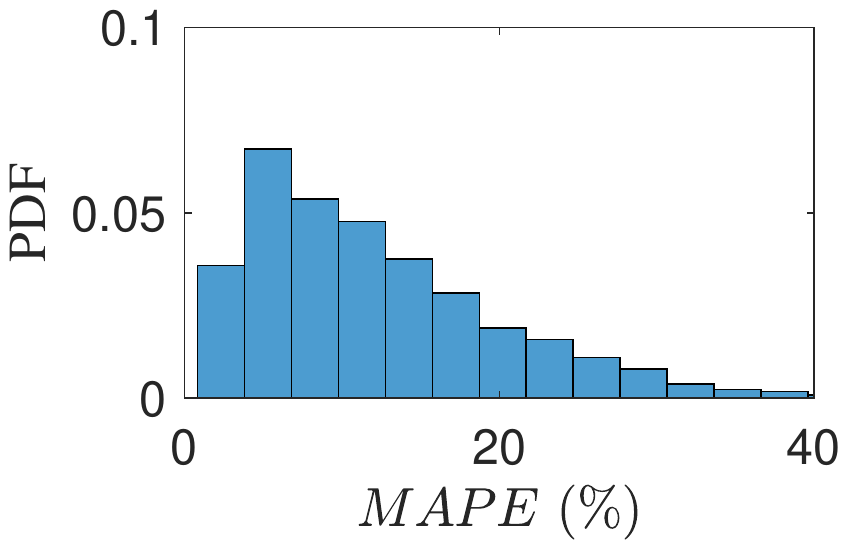}
}
\caption{Empirical distributions of $MAPE$ of disaggregated estimates obtained using the Bi-Modeling method.}
\label{fig:MAPE_distribution_NPY}
\end{figure}

\begin{figure}[htbp]
\centering
\subfloat[PV generation\label{sfig:RMSE_1_NPY}]{
\includegraphics[width=0.37\linewidth]{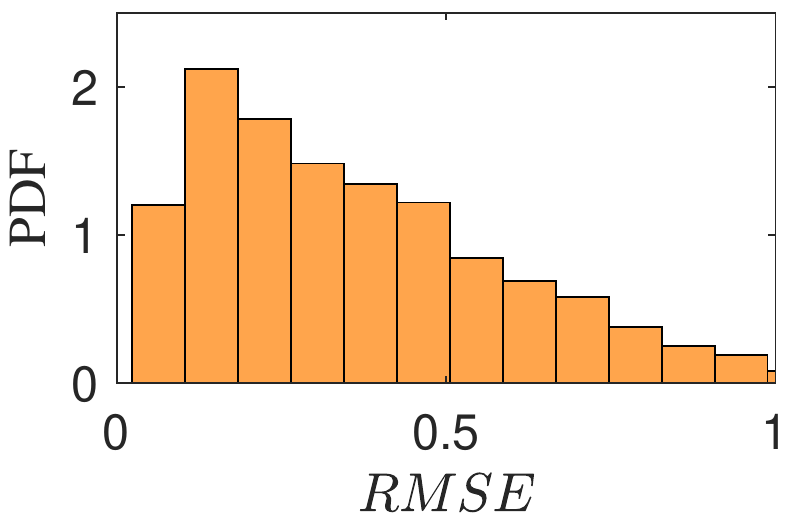}
}
\hfill
\subfloat[Native demand\label{sfig:RMSE_2_NPY}]{
\includegraphics[width=0.38\linewidth]{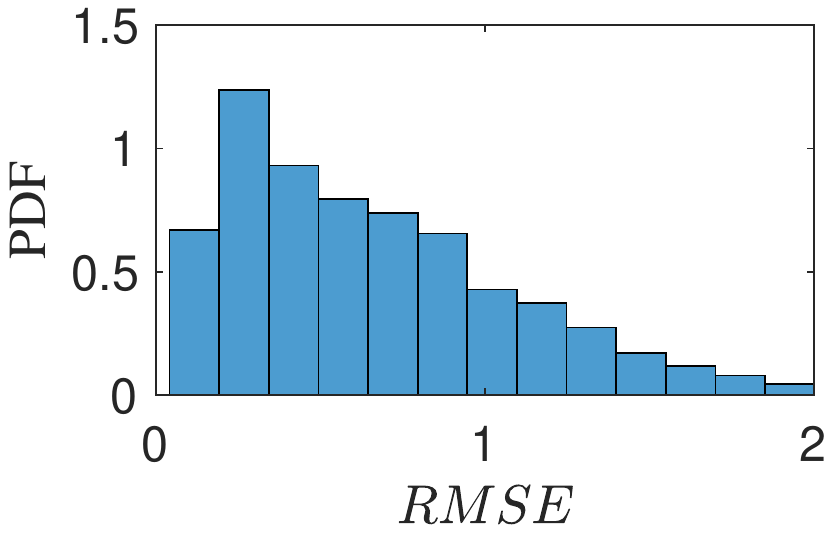}
}
\caption{Empirical distributions of $RMSE$ of disaggregated estimates obtained using the Bi-Modeling method.}
\label{fig:RMSE_distribution_NPY}
\end{figure}

It is vital to compare the performance of our proposed approach with other methods. Since the proposed approach in \cite{nan_peng_yu_2} has been demonstrated to have a relatively better performance than previous methods, we first apply the proposed approach in \cite{nan_peng_yu_2} to conduct PV generation disaggregation using our dataset and then compare its performance with our approach. The approach to be compared is denoted as Bi-Modeling, which employs a statistical model and a physical model to represent the native load and the PV generation, respectively. The Bi-Modeling method utilizes the observable net load series and weather data to optimize model parameters iteratively. A threshold is set to evaluate whether the two models reach a consensus. The results obtained by applying the Bi-Modeling method to our dataset are shown in Fig. \ref{fig:MAPE_distribution_NPY} and \ref{fig:RMSE_distribution_NPY}. It can be seen that our approach has a better performance than the Bi-Modeling method in terms of the $MAPE$ and $RMSE$ of PV generation by comparing Fig.\ref{sfig:MAPE_1_NPY} and \ref{sfig:RMSE_1_NPY} with Fig. \ref{sfig:MAPE_1} and \ref{sfig:RMSE_1}, respectively. In terms of \textit{native demand} disaggregation error comparisons (obtained from Fig. \ref{sfig:MAPE_2}, Fig. \ref{sfig:RMSE_2}, Fig. \ref{sfig:MAPE_2_NPY}, and Fig. \ref{sfig:RMSE_2_NPY}), the results are inconclusive. Further results in terms of \textit{average} $MAPE$ and $RMSE$ are examined as shown in Table \ref{tbl:MAPE_comparison}, and it can be seen that our approach demonstrates smaller disaggregation errors. Note that no single method alone is best in all situations. 

\begin{table}[htbp]
\centering
	\renewcommand{\arraystretch}{1.5}
	\setlength{\tabcolsep}{4.8pt}
	\caption{Average $MAPE$ and $RMSE$ of Estimates}\label{tbl:MAPE_comparison}
	\begin{tabular}{ccc}
	    \toprule[1pt]
		Metrics  & Our Approach &  Bi-Modeling    \\
        Average $MAPE$ of $\hat{G}_{h}$ (\%) & 10.2 & 16.1  \\
        Average $MAPE$ of $\hat{P}_{h}$ (\%) & 9.64  & 12.4 \\
        Average $RMSE$ of $\hat{G}_{h}$      & 0.23 & 0.38  \\
        Average $RMSE$ of $\hat{P}_{h}$      & 0.61  & 0.69 \\
        \bottomrule[1pt]
	\end{tabular}
\end{table}

\section{Application Discussion}\label{sec:discussion}
It is essential to discuss how the disaggregated PV and native demand can be used in practice. These estimates target static applications since the sampling rates of widely available smart meter data are 1-hour, 30-min, or 15-min. To further explain the usefulness of our approach, we primarily focus on three specific applications: 
\subsection{Native Load Monitoring and Forecasting}
Since small-scale rooftop PVs can be disconnected or otherwise absent without prior knowledge, utilities usually adopt a conservative approach in distribution system studies and do not treat small PVs as reliable sources \cite{PV_handbook}. As a result, utilities use the \textit{native} load for conducting conservative scenario analysis instead of the \textit{net} load. Therefore, it is crucial for utilities to monitor the actual native load. In most cases, small-scale rooftop PVs are installed BTM, and only the net load is recorded. Thus, it is necessary to disaggregate the unknown native load and PV generation from the known net load. Our proposed approach can directly provide utilities the estimated native load, which can be further utilized for system operation and design.

The disaggregated estimates can also be used for native load forecasting. As the PV penetration level increases, the native load can be seriously masked by PV generation. Under this condition, it is necessary to separate the native load from the net load first and then perform native load forecasting. For this application, our proposed approach can provide native load estimates to train native load forecasting models.

\subsection{Demand Response}
Due to the existence of BTM PVs, the native demand is masked by PV generation. However, the majority of demand response schemes are designed for native load controlling \cite{kangping_li_solar}. Under this condition, the unknown native demand hinders utilities from applying demand response schemes efficiently because of the invisibility of the real power consumption. Therefore, the native demand of individual customers needs to be separated from the net demand, as our proposed approach fulfills.

\subsection{Service Restoration}
Another application is relevant to service restoration. When restoring cold loads, more power will be drawn by air-conditioning appliances than in normal operation. This power increase is caused by the simultaneous restarting of a large number of appliances and can be several times larger than the normal load. Thus, this abnormal load should be estimated for developing optimal service restoration tactics. One typical way of estimating the abnormal load is to multiply the normal native load before outage by a ratio from a look-up table \cite{PV_handbook,evaluation_2}. To do this, we need to separate the normal native load from the net load. Leveraging the disaggregated native load estimate obtained from our approach can be used in optimizing restoration strategies.

\section{Conclusion}\label{sec:conclusion}
This paper presents a novel robust approach to disaggregate invisible customer-level BTM PV generation and native demand from net demand using smart meter data and solar exemplars. The proposed method employs a limited number of observable solar power exemplars to represent the invisible BTM PV generation. Also, the proposed approach innovatively leverages the significant correlation between nocturnal and diurnal native demands at the timescale of monthly to alleviate the hourly native demand's volatility. In addition, a penalty term is innovatively integrated into the estimation problem to tackle anomalous samples of solar exemplars due to PV failures. The numerical experiments verify that the approach is able to perform disaggregation with excellent accuracy and robustness, which further improves utilities' situational awareness of grid-edge resources. 

The key findings of the paper are summarized as follows: (1) Using real BTM PV generation and native demand data, we have observed that the hourly generation series of a PV can be sufficiently represented using solar power outputs of PVs with similar orientations. In comparison, the hourly customer-level native demand shows higher volatility. (2) Despite the uncertainty of \textit{hourly} native demand, the monthly nocturnal and diurnal native demands are highly correlated. This has inspired us to first estimate the monthly PV generation, then decompose it into hourly solar power. (3) The anomalous data of PV generation is common in practice, and can cause significant disaggregation error. This has motivated us to introduce a penalty term into MLE to reduce the impact of solar exemplars' anomalous samples.

\ifCLASSOPTIONcaptionsoff
  \newpage
\fi



%



\bibliographystyle{IEEEtran}
\bibliography{IEEEabrv,ref}

%








\end{document}